\documentclass [11pt,a4paper]{article}
\usepackage{jcappub}

\usepackage{amsmath,amssymb,amsfonts,amsthm}
\usepackage{bm,bbm}
\usepackage{graphicx}
\usepackage[utf8]{inputenc}
\usepackage[all]{xy}
\usepackage{pifont}
\usepackage{enumitem}

\newcommand{\be}{\begin{equation}}
\newcommand{\ee}{\end{equation}}
\newcommand{\ba}{\begin{eqnarray}}
\newcommand{\ea}{\end{eqnarray}}
\def\bs{\begin{subequations}}
\def\es{\end{subequations}}
\def\a{\alpha}

\def\de{\delta}
\def\g{\gamma}
\def\la{\lambda}

\def\e{\epsilon}

\def\Om{\Omega}
\def\om{\omega}

\def\s{\sigma}

\def\vp{\varphi}

\def\B{\Box}
\def\cA{\mathcal{A}}

\def\cC{\mathcal{C}}

\def\cL{\mathcal{L}}

\def\cN{\mathcal{N}}
\def\cO{\mathcal{O}}
\def\cP{\mathcal{P}}

\def\cR{\mathcal{R}}

\def\cT{\mathcal{T}}

\def\dh{d_{\rm H}}

\def\p{\partial}

\def\H{{\rm H}}
\newcommand{\Eq}[1]{(\ref{#1})}

\def\cob{\color{blue}}

\newcommand{\au}[2]{#1.~#2}
\newcommand{\book}[5]{\emph{#1}, #2, #3, #4 (#5)}
\newcommand{\oarX}[1]{\href{http://arxiv.org/abs/#1}{{\ttfamily\cob #1}}}
\newcommand{\arX}[1]{\href{http://arxiv.org/abs/#1}{{\ttfamily\cob arXiv:#1}}}
\newcommand{\doij}[5]{\href{http://dx.doi.org/#1}{\cob {\it #2} {\bf #3} (#5) #4}}
\newcommand{\doin}[6]{\href{http://dx.doi.org/#1}{\cob {\it #2} {\bf #3 #4} (#6) #5}}
\newcommand{\doinn}[5]{\href{http://dx.doi.org/#1}{\cob {\it #2} {\bf #3} (#5) #4}}

\newcommand{\ndoinn}[5]{\href{#1}{\cob {\it #2} {\bf #3} (#5) #4}}
\newcommand{\proc}[5]{in \emph{#1}, #2  ed., #3, #4 (#5)}
\newcommand{\procsinm}[5]{in \emph{#1}, #2 eds., #3, #4 (#5)}
\newcommand{\tia}[1]{\textit{#1},}

\def\lpl{\ell_{\rm Pl}}
\def\mpl{m_{\rm Pl}}
\def\Mpl{M_{\rm Pl}}
\def\tpl{t_{\rm Pl}}
\def\rme{e}
\def\rmd{d}
\def\rmi{i}

\def\dpl{\delta_{\rm inv}}
\newcommand{\mhyp}{\,\mbox{--}\,}
\renewcommand{\geq}{\geqslant}

\begin{document}

\title{Stochastic gravitational-wave background in quantum gravity}

\author[a]{Gianluca Calcagni,}
\affiliation[a]{Instituto de Estructura de la Materia, CSIC, Serrano 121, 28006 Madrid, Spain}
\emailAdd{g.calcagni@csic.es}

\author[b,c]{Sachiko Kuroyanagi}
\affiliation[b]{Instituto de F\'isica Te\'orica UAM-CSIC, Universidad Aut\'onoma de Madrid, \\calle Nicolás Cabrera 13-15, Cantoblanco, 28049 Madrid, Spain}
\affiliation[c]{Department of Physics, Nagoya University, Chikusa, Nagoya 464-8602, Japan}
\emailAdd{s.kuroyanagi@csic.es}

%\begin{abstract}
\abstract{Among all cosmological quantum-gravity or quantum-gravity-inspired scenarios, only very few predict a blue-tilted primordial tensor spectrum. We explore five of them and check whether they can generate a stochastic gravitational-wave background detectable by present and future interferometers: non-local quantum gravity, string-gas cosmology, new ekpyrotic scenario, Brandenberger--Ho non-commutative inflation and multi-fractional spacetimes. We show that non-local quantum gravity is unobservable, while all the other models can reach the strain sensitivity of DECIGO but not that of LIGO-Virgo-KAGRA, LISA or Einstein Telescope. Other quantum-gravity models with red-tilted spectra (most loop quantum cosmologies) or with exceptionally tiny quantum corrections (Wheeler--DeWitt quantum cosmology) are found to be non-detectable.}
%\end{abstract}

%	02.40.Gh	Noncommutative geometry
% 04.60.-m	Quantum gravity
% 05.45.Df	Fractals
% 11.10.Kk	Field theories in dimensions other than four
% 11.10.Lm	Nonlinear or nonlocal theories and models
% 11.10.Nx	Noncommutative field theory

%%\pacs{05.45.Df,11.10.Kk,11.10.Lm,04.60.-m,02.50.-r,02.30.Cj,02.40.Gh,11.10.Nx}
%\preprint{JHEP03(2010)120 \hspace{8.3cm} arXiv:1001.0571}
%\keywords{modified gravity, alternatives to inflation, transplanckian physics, cosmology of theories beyond the SM}

%\begin{document}
\maketitle

%%%%%%%%%%%%%%%%%%%%%%%%%%%%%%%%%%%%%%%%%%%%%%%%%%%%%%%%%%%%%%%%%%%%%%%%%%%%%%%%%%%%%%%%%%%%%%%%%%%%
%%%%%%%%%%%%%%%%%%%%%%%%%%%%%%%%%%%%%%%%%%%%%%%%%%%%%%%%%%%%%%%%%%%%%%%%%%%%%%%%%%%%%%%%%%%%%%%%%%%%

\section{Introduction}

The rapidly expanding field of gravitational-wave (GW) astronomy has become an excellent testing ground for old and new ideas about gravity, from general relativity itself to modified gravity models and advanced theories of quantum gravity. Although available observations fully confirmed general relativity \cite{TheLIGOScientific:2016src,LIGOScientific:2020stg}, there is plenty of room for exploring alternative theories. In particular, quantum gravity can alter the production \cite{Yunes:2016jcc,Kobakhidze:2016cqh,BYY,TaYa,Mas18,Giddings:2019ujs,Agullo:2020hxe} and the propagation \cite{Yunes:2016jcc,EMNan,qGW,ArCa2,Wang:2017igw,Calcagni:2019kzo,Belgacem:2019pkk,Calcagni:2019ngc,Wang:2020pgu,Garcia-Chung:2020zyq} of astrophysical GWs or appear in the strain noise of interferometers as an anomalous signal \cite{Calcagni:2019ngc,Ame98,NgDa2,Ame13,Bosso:2018ckz,ACCR,AMY}. In this paper, we consider primordial GWs in quantum gravity and their remnant as a stochastic GW background (SGWB). Since there exist several early universe models in quantum gravity \cite{CQC}, a natural question is whether some of them would be able to leave a characteristic imprint in the SGWB within the sensitivity range of present or future interferometers such as LIGO-Virgo-KAGRA (LVK) \cite{TheLIGOScientific:2016wyq,Abbott:2017xzg,Akutsu:2018axf}, LISA \cite{Bartolo:2016ami,Caprini:2019pxz}, Einstein Telescope (ET) \cite{Maggiore:2019uih} and DECIGO \cite{Seto:2001qf,Kawamura:2011zz,Kawamura:2020pcg}. This possibility would become substantial for any model predicting a blue-tilted tensor spectrum (tensor spectral index $n_{\rm t}>0$) with a sufficiently high tensor amplitude (high tensor-to-scalar ratio $r$). 

As we will see, it turns out that only very few quantum-gravity models have a blue tilt and most of them fail to produce a large signal since $r$ is very small in the region of the parameter space where $n_{\rm t}>0$ at high frequencies. Four of these models (string-gas cosmology, new ekpyrotic scenario, Brandenberger--Ho non-commutative inflation, multi-fractional inflation) reach the sensitivity threshold of DECIGO.
%and also of ET in the case of string-gas cosmology. These positive results give us hope that GW astronomy could, indeed, say something about fundamental extensions of Einstein's theory. 
On the other hand, the model of inflation in non-local quantum gravity reduces to ordinary Starobinsky gravity at large frequencies and the spectrum is not lifted. This result is non-trivial because it relies on the full expression of the tensor spectrum rather than on it slow-roll parametrization, which we show to break down due to large non-local corrections.

The five models we will concentrate on are:
\begin{enumerate}
\item \emph{Non-local Starobinsky inflation} \cite{Craps:2014wga,Koshelev:2016xqb,Koshelev:2017tvv,SravanKumar:2018dlo,Koshelev:2020foq}. This is a cosmological model arising directly in non-local quantum gravity \cite{Kra87,Kuz89,Tom97,Modesto:2011kw,BGKM,BCKM,MoRa1,MoRa2,Buoninfante:2018mre} (see \cite{Modesto:2017sdr} for a review), where the bare gravitational action is second-order in curvature tensors and is endowed with non-local operators, called form factors, with infinitely many derivatives. Second-order curvature terms make the ensuing perturbative field theory renormalizable (as in local Stelle gravity \cite{Ste77,Ste78,Julve:1978xn,Fradkin:1981iu,Fradkin:1981hx,Avramidi:1985ki,HOW}), while non-local operators preserve unitarity. In the local limit expanding the form factors \cite{Briscese:2013lna}, the theory reduces to Starobinsky gravity \cite{Starobinsky:1980te,Mukhanov:1981xt,Barrow:1983rx,Starobinsky:1983zz,Whitt:1984pd,Kofman:1985aw,Starobinsky:1987zz,Maeda:1987xf,Maeda:1988rb} and, therefore, it can sustain a viable inflationary era \cite{Akrami:2018odb}. However, if one retains the full non-local form factors, one obtains a non-local version of Starobinsky gravity with different, characteristic inflationary observables in the tensor sector, while the scalar spectrum is unchanged \cite{Koshelev:2016xqb}. As we will see below, however, while the tensor index can be positive depending on the choice of form factor, this observable is not sufficient to determine correctly the high-frequency behaviour of the tensor spectrum because higher-order observables get non-negligible non-local corrections. When the full spectrum is considered, the blue tilt found from the tensor index and its running turns out to be spurious and the spectrum remains below the detection threshold of DECIGO and the other experiments. 
\item \emph{String-gas cosmology} \cite{BrVa1,BEK,NBVa,BrNPV,BNPV,Brandenberger:2008nx,BrNP,Bra11,Bra12,Brandenberger:2015kga,Kamali:2020drm,Bernardo:2020nol,Bernardo:2020bpa}. This model, stemming from string theory, produces primordial spectra via a thermal mechanism alternative to inflation. In a compact space, the excitation modes of a thermal ensemble of strings are momentum modes and winding modes. The energy of winding modes decreases with the size of the available space and their number increases with the energy, so that, for an adiabatic process, winding modes dominate the thermal bath in a small space. The temperature of this bath cannot rise indefinitely and reaches a maximal temperature $T_{\rm H}$ called Hagedorn temperature \cite{Hag65}. Since scales smaller than the string length or energies higher than $T_{\rm H}$ are not physically reachable, this mechanism based on string thermodynamics can solve the big-bang problem and also offer an alternative to inflation solving the horizon problem \cite{BEK}. The universe starts with an almost constant scale factor and a temperature slightly lower than $T_{\rm H}$. Size \cite{Patil:2004zp,Patil:2005fi} and shape \cite{Brandenberger:2005bd} moduli and the dilaton \cite{Danos:2008pv} are stabilized. Both scalar and tensor spectra are generated thermally. In particular, tensor modes are generated by anisotropic pressure terms in the energy-momentum tensor, but near the Hagedorn temperature the thermal bath is dominated by winding modes and the pressure decreases. Thus, there is a decrease of power at low $k$ and a slight blue tilt. Eventually, winding modes decay, three spatial directions open up and the others stay compact, leading to a radiation-dominated era.
\item \emph{New ekpyrotic scenario} \cite{Brandenberger:2020tcr,Brandenberger:2020eyf,Brandenberger:2020wha}. At the density of the string scale, new degrees of freedom govern the effective four-dimensional cosmological dynamics. At the time when such density is reached, the dynamics is dominated by an S-brane, a space-like hypersurface with zero energy density and negative pressure that induces a transition between a contracting phase (ekpyrosis) and an expanding one. The inhomogeneous perturbations generated before the bounce become almost scale-invariant after passing through it. The amplitude of the tensor spectrum depends on the ratio of the string scale to the Planck scale and is enhanced at super-horizon scales. If the S-brane has zero shear, then $r\lesssim 10^{-3}$, the scalar spectrum is red-tilted and the tensor spectrum is blue-tilted \cite{Brandenberger:2020eyf}.
\item \emph{Brandenberger--Ho non-commutative inflation} \cite{BH,HL1,Fukuma,TMB,HL2,HL3,KLM,KLLM,cai04,Cal4,Cal5,CT04,Calcagni:2013lya}. In this scenario, time and space coordinates do not commute and, as a consequence, the action of the inflaton scalar field driving the early phase of acceleration is decorated with *-products. This structure alters the scalar and tensor primordial spectra in a way that depends on whether the perturbation modes have a wave-length smaller (ultraviolet limit, UV) or larger (infrared limit, IR) than the fundamental length scale appearing in the commutation algebra. In the particular case of natural inflation \cite{FFO,ABFFO,KNPe}, the IR limit of this model has a blue-tilted tensor spectrum and is compatible with \textsc{Planck} data \cite{Calcagni:2013lya}. Here the time-momentum uncertainty induces a $k$-dependence in the effective mass term of the GW propagation equation. This dependence is inherited by the tensor spectrum and, for a certain choice of parameters, it leads to an enhancement at small scales.
\item \emph{Multi-fractional inflation} \cite{frc14}. It arises in multi-fractional spacetimes \cite{revmu}, theories which realize explicitly dimensional flow, a universal feature of quantum gravity such that the dimension of spacetime changes with the probed scale \cite{tH93,Car09,fra1,Car17}. Integrals and derivatives acquire an anomalous multi-scaling, the action gets a new discrete symmetry at short scales and standard cosmology is modified accordingly \cite{frc14,frc11,Calcagni:2017via}. In particular, the scalar and tensor spectrum are modified in such a way that the slow-roll approximation can be relaxed. Here we will consider a previously ignored corner in the parameter space that produces a blue-tilted tensor spectrum. The mechanism to achieve this is purely geometrical: the dimension of spacetime is smaller at small scales/large frequencies but the density of states is the same at all scales, so that small-scale modes have ``less spacetime'' available and their number increases, hence an enhancement of the power spectrum at large frequencies.
\end{enumerate}
The advantages and disadvantages of these models are listed in Tab.\ \ref{tab1}.
\begin{table}
\makebox[\textwidth][c]{{\renewcommand{\arraystretch}{1.2}
{\footnotesize\begin{tabular}{lll}\hline
																& {\bf Advantages} & {\bf Disadvantages}	\\\hline\hline
\begin{minipage}[t]{0.25\textwidth}{\bf Non-local Starobinsky}\\ {\bf inflation}\end{minipage} &  \begin{minipage}[t]{0.35\textwidth}\begin{itemize}[leftmargin=*]\item Embedded in a full quantum gravity. \item Easily understandable physical setting and cosmology.\item Acceleration from curvature.\end{itemize}\end{minipage} &  \begin{minipage}[t]{0.35\textwidth}\begin{itemize}[leftmargin=*]\item Non-uniqueness of form factors, especially on curved backgrounds. \item Unclear UV properties on curved backgrounds. \item Difficult to extract phenomenology.\end{itemize}\vspace{.05cm}\end{minipage} \\\hline
\begin{minipage}[t]{0.25\textwidth}{\bf String-gas cosmology}\end{minipage}       &  \begin{minipage}[t]{0.35\textwidth}\begin{itemize}[leftmargin=*]\item Embeddable in string theory.\item Few free parameters.\item Does not require inflaton.\end{itemize}\end{minipage} &  \begin{minipage}[t]{0.35\textwidth}\begin{itemize}[leftmargin=*]\item Analytic temperature dependence $T(k)$ on wave-number unknown.\item Requires bounce or short inflation to solve flatness problem.\item Early-universe observables not following a slow-roll hierarchy.\end{itemize}\vspace{.05cm}\end{minipage} \\\hline
\begin{minipage}[t]{0.25\textwidth}{\bf New ekpyrotic}\\ {\bf scenario}\end{minipage}       &  \begin{minipage}[t]{0.35\textwidth}\begin{itemize}[leftmargin=*]\item Embeddable in string theory.\item Does not require inflaton.\item Simple consistency relations.\end{itemize}\end{minipage} &  \begin{minipage}[t]{0.35\textwidth}\begin{itemize}[leftmargin=*]\item Dilaton and moduli stabilization not under full control.\item Complicated bouncing dynamics.\item Early-universe observables not following a slow-roll hierarchy.\end{itemize}\vspace{.05cm}\end{minipage} \\\hline
\begin{minipage}[t]{0.25\textwidth}{\bf Non-commutative}\\ {\bf inflation}\end{minipage}       &  \begin{minipage}[t]{0.35\textwidth}\begin{itemize}[leftmargin=*]\item Realizes fuzziness (frequent or universal feature in quantum gravity). \item Few free parameters, easily falsifiable.\end{itemize}\end{minipage} &  \begin{minipage}[t]{0.35\textwidth}\begin{itemize}[leftmargin=*]\item Not embedded in a full quantum gravity.\item Large non-commutative effects speculative.\item Requires inflaton with selected potential.\end{itemize}\vspace{.05cm}\end{minipage} \\\hline
\begin{minipage}[t]{0.25\textwidth}{\bf Multi-fractional}\\ {\bf inflation}\end{minipage}      &  \begin{minipage}[t]{0.35\textwidth}\begin{itemize}[leftmargin=*]\item Realizes dimensional flow (universal feature in quantum gravity). \item Very characteristic predictions, easily falsifiable. \item Easy analytic treatment.\item Does not require special inflaton potentials.\end{itemize}\vspace{.05cm}\end{minipage} &  \begin{minipage}[t]{0.35\textwidth}\begin{itemize}[leftmargin=*]\item Three out of four theories are not renormalizable quantum gravities. \item Bizarre spacetime geometries when $n_{\rm t}>0$.\item Requires inflaton.\end{itemize}\end{minipage} \\\hline
\end{tabular}}}}
\caption{\label{tab1} Pros and cons of the main models considered in this paper.}
\end{table}

A general positive feature we will discover in this paper is that none of these models requires a fine tuning of the parameters in order to reach the sensitivity threshold of interferometers. However, most of them do not go beyond the level of DECIGO. 
%The only exception is string-gas cosmology, which reaches the sensitivity of ET (and higher).
%, and non-local quantum gravity, where a non-trivial running of the tensor spectral index is crucial to boost the spectrum up to LIGO-Virgo-KAGRA sensitivity (and higher).

In section \ref{sec2}, the definitions and experimental values of the main cosmological observables considered in this paper will be reviewed. In section \ref{sec3}, we summarize models of quantum gravity whose primordial spectra have been investigated in the literature and that, as it turns out, give rise to an undetectable SGWB signal for three possible reasons: (i) the tensor spectrum is red tilted; (ii) the tensor spectrum is blue-tilted but only minimally so; (iii) the tensor spectrum is strongly blue-tilted but the tensor-to-scalar ratio is too small. In section \ref{sec4}, we describe the above models with blue tilt and we compare their SGWB spectrum with the sensitivity curves of LIGO-Virgo-KAGRA, LISA, Einstein Telescope and DECIGO. Bounds from big-bang nucleosynthesis (BBN) are considered in section \ref{sec5}, where we will 
%determine a new bound from BBN on the tensor index with or without tensor running (section \ref{bbnbo}). 
discuss the impact of including the tensor running and constraints on the models.
Section \ref{sec6} is devoted to a discussion of these results.

%%%%%%%%%%%%%%%%%%%%%%%%%%%%%%%%%%%%%%%%%%%%%%%%%%%%%%%%%%%%%%%%%%%%%%%%%%%%%%%%%%%%%%%%%%%%%%%%%%%%
%%%%%%%%%%%%%%%%%%%%%%%%%%%%%%%%%%%%%%%%%%%%%%%%%%%%%%%%%%%%%%%%%%%%%%%%%%%%%%%%%%%%%%%%%%%%%%%%%%%%

\section{Cosmological observables}\label{sec2}

%%%%%%%%%%%%%%%%%%%%%%%%%%%%%%%%%%%%%%%%%%%%%%%%%%%%%%%%%%%%%%%%%%%%%%%%%%%%%%%%%%%%%%%%%%%%%%%%%%%%

\subsection{CMB observables}

We collect some definitions and cosmological formul\ae\ of common use. On a Friedmann--Lema\^itre--Robertson--Walker (FLRW) background with scale factor $a(t)$, the number of e-foldings from the time $t$ of horizon crossing until the end of inflation at $t_{\rm e}$ is 
\be\label{defiN}
\cN:=\ln \frac{a(t_{\rm e})}{a(t)} =\int_t^{t_{\rm e}}\rmd t'\,H(t')\,.
\ee
In particular, $\dot\cN=-H$, where $H:=\dot a/a$ is the Hubble parameter. The first two slow-roll parameters in terms of the Hubble expansion are
\be
\e :=-\frac{\dot H}{H^2}=\frac{\rmd\ln H}{\rmd \cN}\,,\qquad
\eta := -\frac{\ddot H}{2H\dot H} = \e+\frac12\frac{\rmd\ln\e}{\rmd \cN}\,.\label{epseta}
\ee
The time derivative of $\e$ is second-order in the slow-roll parameters, 
\be\label{dote}
\dot\e=2H\e(\e-\eta)\,.
\ee
Defining $\e_V:=(\Mpl^2/2)(V_{,\phi}/V)^2$ as the first slow-roll parameter in terms of the inflaton potential $V(\phi)$, where $\Mpl=(8\pi)^{-1/2}\mpl$ $=(8\pi G)^{-1/2}\approx2.44\times 10^{18}\,{\rm GeV}$ is the reduced Planck mass, in the slow-roll approximation $\e\simeq\e_V$ in standard cosmology, but this relation can change in quantum gravity; an example is given in \cite{BCT2}. Also, in standard cosmology (but not in models with modified Friedmann equations) $\eta=\eta_\phi:= -\ddot{\phi}/(H \dot\phi)$ and the combination \Eq{dote} is usually positive for the most common types of inflationary potential \cite{CQC}. For instance, for a monomial potential $V\propto\phi^p$ (large-field models), one has $\e=p/(4\cN+p)$, $\eta=(p-2)/(4\cN+p)$ and $\dot\e>0$; for small-field models with $V\propto 1-(\phi/\phi_0)^p$ as well as for natural inflation \Eq{napotential}, one has $\eta<0$, $|\eta|\gg\e$ and $\dot\e>0$. In other words, in Einstein gravity the first slow-roll parameter $\e$ typically grows in time until inflation ends at $\e=1$.

Let $\cP_{\rm s}(k)$ and $\cP_{\rm t}(k)$ be, respectively, the primordial scalar and tensor spectrum as a function of the spatial comoving wave-number $k=|{\bm k}|$. An observable of interest is the tensor-to-scalar ratio
\be\label{ar}
r := \frac{\cP_{\rm t}}{\cP_{\rm s}}\,.
\ee
These quantities are evaluated at horizon crossing, i.e., whenever the Hubble parameter appears it should be replaced by $H=k/a$. Consequently, derivatives with respect to the wave-number can be translated into time or e-folds derivatives and vice versa:
\be\label{tk}
\frac{\rmd}{\rmd\ln k}=\frac{1}{H(1-\e)}\frac{\rmd}{\rmd t}\simeq \frac{1}{H}\frac{\rmd}{\rmd t}= -\frac{\rmd}{\rmd\cN}\,.
\ee
In particular, the tensor spectral index and its running are defined by
\be
n_{\rm t} := \frac{\rmd\ln\cP_{\rm t}}{\rmd\ln k}\,,\qquad
\a_{\rm t} := \frac{\rmd n_{\rm t}}{\rmd\ln k}\,.\label{ntat}
\ee
In the scalar sector, these observables are replaced by the scalar spectral index $n_{\rm s}-1$ and the scalar running $\a_{\rm s}$. In standard cosmology, 
\be\label{eingrav}
n_{\rm t}\propto -\e<0\,,\qquad \a_{\rm t}\propto-\dot\e<0\qquad \textrm{(Einstein gravity)}\,,
\ee
for what commented below \Eq{dote}: the tensor spectrum is red-tilted and it becomes redder at higher $k$. In quantum gravity, the sign of both observables can flip, as we will see in due course.

%In the models of non-local quantum gravity and multi-fractional inflation,
To compute the SGWB, we will use the direct expression $\cP_{\rm t}(k)$ of the primordial tensor spectrum or, wherever possible, the standard parametric form
\be\label{Ptk}
\cP_{\rm t}(k)=\cP_{\rm t}(k_0)\,\exp\left[n_{\rm t}(k_0)\,\ln\frac{k}{k_0}+\frac{\a_{\rm t}(k_0)}{2}\left(\ln\frac{k}{k_0}\right)^2\right],\qquad \cP_{\rm t}(k_0)=r(k_0)\,\cP_{\rm s}(k_0)\,,
\ee
where $r$, $n_{\rm t}$ and $\a_{\rm t}$ are given by the theory and are calculated at the pivot scale $k_0$, while $\cP_{\rm s}(k_0)$ is the measured amplitude of the comoving curvature perturbation. For multi-fractional inflation, we will use a different parametrization.

In some cases, the factor responsible for the enhancement of the SGWB at interferometer scales is the tensor running. During inflation, the running term does not play an important role because $\a_{\rm t}\ll 1$ and $[\ln(k/k_0)]^2\ll 1$. However, at higher frequencies $[\ln(k/k_0)]^2$ increases and the running term can be large enough to inflect the spectrum upwards towards the strain sensitivity range of near-future experiments. This effect is consistent with the parametrization \Eq{Ptk} as long as $\a_{\rm t}$ is small, which is the case in the models considered here with the notable exceptions of non-local gravity and string-gas cosmology.

%In the case of string-gas cosmology, this approximation will be used because the exact $k$-dependence in $\cP_{\rm t}$ and the other observables is not known but there is a consistency relation between $n_{\rm t}$ and $n_{\rm s}$, hence between $\a_{\rm t}$ and $n_{\rm s}$. Using the observational constraints on $n_{\rm s}$ and $\a_{\rm s}$, one can find the corresponding values of $n_{\rm t}$ and $\a_{\rm t}$ and plot the spectrum \Eq{Ptk}. In the case of non-commutative inflation, $n_{\rm t}(\cN)$ and $\a_{\rm t}(\cN)$ are calculated from a background scalar-field inflationary model and then plugged into \Eq{Ptk}.

We use the values of cosmological parameters provided by the \textsc{Planck} Legacy release \cite{Akrami:2018odb} from the \textsc{Planck}+TT+TE+EE+lowE+lensing+BK15+BAO data set at the pivot scale %$k_0=0.05\,{\rm Mpc}^{-1}$,
\be\label{pivot}
k_0=0.05\,{\rm Mpc}^{-1}\,,
\ee
where one has
\be\label{psbo}
\cP_{\rm s}(k_0)=2.0989\times10^{-9}\,,\qquad
h=0.6736 \,,\qquad
\Omega_{\rm m}=0.3153\,.
\ee
For inflationary observables, the following constraints on $r$, $n_{\rm s}$ and $\a_{\rm s}$ were obtained for a $\Lambda$CDM+$r$+$\a_{\rm s}$ model:
\bs\label{obspl}\ba
&r<0.068\qquad\qquad\qquad\quad\,\, &\textrm{(95\% CL)}\,,\label{boundr}\\
&n_{\rm s}=0.9658\pm 0.0040 \qquad\,\,\,\, &\textrm{(68\% CL)}\,,\label{obsns}\\
&\a_{\rm s}=-0.0066\pm0.0070\qquad &\textrm{(68\% CL)}\,,
\ea\es
where CL is the confidence level. There are no strong model-independent cosmic microwave background (CMB) constraints on $n_{\rm t}$ and $\a_{\rm t}$ but model-dependent consistency relations between $r$ and $n_{\rm t}$ can place an indirect bound on the tensor index.

%%%%%%%%%%%%%%%%%%%%%%%%%%%%%%%%%%%%%%%%%%%%%%%%%%%%%%%%%%%%%%%%%%%%%%%%%%%%%%%%%%%%%%%%%%%%%%%%%%%%

\subsection{GW observables}

A gravitational wave is described as a small dimensionless tensor perturbation around a flat homogeneous cosmological background, 
\be
\rmd s^2 = a^2(\tau)\left\{ - \rmd\tau^2 + \left[\delta_{ij} + h_{ij} (\tau,{\bm x})\right] \rmd x^i \rmd x^j \right\}\,.
\ee
Here we use the conformal time $\tau$, related to cosmic time $t$ by $\rmd\tau = \rmd t / a(t)$. It is convenient to Fourier transform $h_{ij} (\tau,{\bm x})$ as
\be
%\label{ }
h_{ij} (\tau, {\bm x}) = \sum_{l = +, \times} \int \frac{\rmd^3 {\bm k}}{(2\pi)^{3/2}}\,\epsilon_{ij}^l\, h_{\bm k}^l (\tau)\, \rme^{\rmi {\bm k}\cdot {\bm x}}\,,
\ee
where the polarization tensor $\epsilon_{ij}^l$ satisfies the symmetric, transverse-traceless condition and is normalized by the relation $\sum_{i,j} \epsilon_{ij}^l (\epsilon_{ij}^{l'})^\ast = 2 \delta^{ll'}$.
	
The energy density $\rho_\textsc{gw}$ of GWs in general relativity is the spatial average of the kinetic energy of the perturbation:
\be
\rho_\textsc{gw}:=\frac{\Mpl^2}{8a^2} \left\langle  \left( \partial_\tau h_{ij}  \right)^2+  \left( \nabla h_{ij}  \right)^2 \right\rangle
= \frac{\Mpl^2}{4}  \int \rmd \ln k \left( \frac{k}{a} \right)^2 \frac{k^3}{\pi^2} \sum_l \left| h_{\bm k}^l \right|^2\,. 
\ee
The amplitude of the SGWB is commonly characterized by the dimensionless density parameter 
\be\label{Omgw0}
\Om_\textsc{gw}(k, \tau) :=  \frac{1}{\rho_{\rm crit}} \frac{\rmd\rho_\textsc{gw}}{\rmd \ln k}\,,
\ee
where $\rho_{\rm crit}:=3\Mpl^2H_0^2$ is the critical energy density. The tensor power spectrum is defined as 
\be
%\label{ }
\Delta^2_t(k, \tau) := \frac{\rmd\langle 0|h_{ij}^2|0\rangle}{\rmd\ln k} = \frac{k^3}{\pi^2} \sum_l \left| h^l_{{\bm k}} \right|^2 =: \mathcal{P}_{\rm t} (k)\,\cT^2 (k, \tau)\,,
\ee
where $\mathcal{P}_{\rm t} (k)$ is the primordial tensor power spectrum and $\cT(k, \tau)$ is the transfer function describing the deformation of the primordial spectrum by the evolution of GWs after horizon crossing. Therefore, the energy density today can be written as 
\be\label{Omgw}
\Om_\textsc{gw}(k, \tau_0) =\frac{k^2}{12a_0^2H_0^2}\cP_{\rm t}(k)\,\cT^2(k, \tau_0)\,,
\ee
where the subscript 0 indicates quantities evaluated at the present time.

The shape of the transfer function strongly depends on the Hubble expansion history of the universe; for details, see \cite{Kuroyanagi:2008ye}. In the case of the instant reheating scenario, where soon after inflation the universe enters a radiation-dominated phase, the transfer function for the SGWB amplitude today is \cite{Nakayama:2008wy}
\be
\label{Tk}
\cT^2 (k, \tau_0) =  
\Omega_{\rm m}^2 \left[ \frac{g_\ast ( T_{\rm in})}{g_{\ast 0}} \right]
 \left[ \frac{g_{\ast s0}}{g_{\ast s} (T_{\rm in})} \right]^{4/3}
 \left[ \frac{ 3 j_1 (k\tau_0)}{k \tau_0} \right]^2 T_{\rm eq}^2 (k) \,.
\ee 
The first spherical Bessel function $j_1 (k \tau_0)$ can be approximated as $ j_1 (k \tau_0) \simeq 1 / (\sqrt{2} k \tau_0)$ in the limit $k \tau_0 \rightarrow 0$. The values of the relativistic degrees of freedom $g_\ast ( T_{\rm in})$ and of its counterpart for entropy $g_{\ast s} (T_{\rm in})$ change depending on the cosmic temperature at which the mode $k$ enters the horizon, and it changes the spectral shape \cite{Watanabe:2006qe}. To incorporate this effect in the GW spectrum, we introduce the fitting function \cite{Kuroyanagi:2014nba}
\begin{equation*}
g_\ast [T_{\rm in}(k)]=g_{\ast 0} 
\frac{A_1+\tanh \left[-2.5\log_{10}\left( \frac{k/2\pi}{2.5\times 10^{-12} ~{\rm Hz}}\right) \right]}{A_1+1}\,
\frac{A_2+\tanh \left[-2.0\log_{10}\left( \frac{k/2\pi}{6.0\times 10^{-9} ~{\rm Hz}} \right) \right]}{A_2+1}
\end{equation*} 
where $A_1=(-1-10.75 /g_{\ast 0})/(-1+10.75/g_{\ast 0})$ and $A_2=(-1-g_{\rm max}/10.75) /(-1+g_{\rm max}/10.75)$.  For $g_{\rm max}$, we assume the sum of the Standard-Model particles, $g_{\rm max}=106.75$.  The same formula can be used for the counterpart for entropy $g_{\ast s}(T_{\rm in})$ by replacing $g_{\ast 0}=3.36$ with $g_{\ast s0}=3.91$. The other fitting function $T_{\rm eq}^2(k)$ is given by \cite{Turner:1993vb}
\be
T_{\rm eq}^2(k) = 1 + 1.57 x_{\rm eq} + 3.42 x_{\rm eq}^2,\qquad x_{\rm eq} :=  \frac{k}{k_{\rm eq}}\,,
\ee
where
\be
k_{\rm eq} = 7.1 \times 10^{-2}\, \Omega_{\rm m} h^2~{\rm Mpc}^{-1} 
\ee
is the comoving wave-number at radiation-matter equivalence, corresponding to 
$f_{\rm eq}\approx 9.9\times 10^{-17} (\Omega_{\rm m} h^2/0.143) \,{\rm Hz}$.

%%%%%%%%%%%%%%%%%%%%%%%%%%%%%%%%%%%%%%%%%%%%%%%%%%%%%%%%%%%%%%%%%%%%%%%%%%%%%%%%%%%%%%%%%%%%%%%%%%%%
%%%%%%%%%%%%%%%%%%%%%%%%%%%%%%%%%%%%%%%%%%%%%%%%%%%%%%%%%%%%%%%%%%%%%%%%%%%%%%%%%%%%%%%%%%%%%%%%%%%%

\section{Discarding models}\label{sec3}

Obtaining a primordial blue-tilted tensor spectrum in quantum gravity is difficult. Despite the abundance of viable cosmological inflationary models in quantum gravity, a close scrutiny reveals that most of them predict a red tilt and those that have a blue tilt often lead to unobservable effects because $n_{\rm t}$ or $r$, or both, are too close to zero. 

For instance, the large class of flux-compactification models in string cosmology is uniformly characterized by $n_{\rm t}<0$ \cite{CQC,Baumann:2014nda}. A case apart is the old ekpyrotic scenario, which predicts a strongly blue-tilted tensor index $n_{\rm t}=2$ \cite{KOST1}. While early versions of the model are ruled out because they have also a blue-tilted scalar spectrum, in a recent single-field version perturbations are generated before the ekpyrotic phase and the scalar spectrum is safely red-tilted \cite{KhSt1,KhSt2}. However, in all the realizations of the model the tensor-to-scalar ratio is exceptionally small and the resulting SGWB is well below the detection threshold of any present or future interferometer \cite{Boyle:2003km}.

Similarly, we call the Hamiltonian approaches to quantum gravity out of the game. In Wheeler--DeWitt canonical quantum cosmology, the semi-classical limit of the Wheeler--DeWitt equation for the wave-function of the Universe admits two solutions, one which is continuous along the imaginary axis and the other exhibiting a jump at the origin \cite{Kiefer:2011cc,Bini:2013fea}. The continuous solution leads to red-tilted spectra in both the scalar and the tensor sector \cite{Brizuela:2015tzl,Brizuela:2016gnz,Brizuela:2019jzv}. The discontinuous solution has been investigated less but the results in the scalar sector \cite{Bini:2013fea,ADPre} indicate an enhancement of power at small scales, i.e., a blue tilt. The quantum correction to the power spectrum is so strongly suppressed during inflation that it plays no role at late times either.

The tensor spectrum of Wheeler--DeWitt quantum cosmology reads
\be\label{wdw} 
\cP_{\rm t}(k)\simeq\cP_{\rm t}^{(0)}(k) \left[1\pm c\,(\lpl H)^2\left(\frac{k_0}{k}\right)^3\right],
\ee
where $\cP_{\rm t}^{(0)}(k)\propto H^2$ is the standard spectrum at horizon crossing ($k=aH$), $c>0$ is a known numerical constant, $k_0$ is the pivot scale of the experiment (for the CMB, typically $k_0=0.05\,{\rm Mpc}^{-1}$ or $k_0=0.002\,{\rm Mpc}^{-1}$) and
\be\label{lpmpc}
\lpl\approx 10^{-35}\,{\rm m}=5\times 10^{-58}\,{\rm Mpc}
\ee
is the Planck length. The case with blue (respectively, red) tilt corresponds to the $-$ ($+$) sign. The same corrections to the inflationary spectra have been obtained in a similar approach \cite{Kamenshchik:2014kpa,Kamenshchik:2015gua,Kamenshchik:2016mwj}. The strong suppression of the $(\lpl H)^2\ll 1$ term is further increased at late times by the $(k_0/k)^3$ factor, since at the frequencies of LISA and DECIGO $k_0/k\sim 10^{-15}\mhyp 10^{-13}$.

Loop quantum cosmology is another Hamiltonian framework, based on the quantization of gravity in Ashtekar--Barbero variables. There are three main approaches to cosmological perturbations within loop quantum cosmology.
\begin{itemize}
\item In the dressed-metric approach \cite{AAN2,AAN3}, the tensor spectrum is red tilted \cite{Agullo:2015tca,Li:2019qzr}. 
\item In the effective-constraints approach \cite{BBCGK}, one can consider quantum corrections coming from inverse-volume operators, from holonomies or, more realistically, from both. If one considers only inverse-volume corrections, the inflationary spectra are compatible with observations \cite{BCT2,Zhu15} but the tensor spectrum is red tilted, since $n_{\rm t} \simeq -2\e_V-\dpl$, where $\dpl(k)>0$ is a positive inverse-volume quantum correction \cite{BCT2}. On the other hand, the case with only holonomy corrections predicts a blue-tilted tensor spectrum but it is ruled out observationally \cite{BoBGS}. To the best of our knowledge, the case with both types of corrections has not been explored yet.
\item In the hybrid-quantization approach \cite{FMO1,CGMM,deBO}, the value and sign of the spectral index $n_{\rm t}(k)$ depend on the background effective solution and, even more importantly, on the vacuum on which to perturb such background. The $k$-dependence of the tensor index can be found via a numerical analysis and it turns out that for some choices of vacuum $n_{\rm t}<0$, while for others the spectrum oscillates rapidly and a blue tilt can be generated at certain frequencies \cite{deBO,Gomar:2017yww}. However, what matters for the formation of a SGWB is the average trend of the spectrum and, in all these cases, it decreases in $k$ when $k$ is sufficiently large. Therefore, it is unlikely that this model could generate a detectable SGWB, for any vacuum choice. A detailed numerical study, which we will not pursue here, could give a more precise answer.
\end{itemize}
There are other quantum-gravity models where the tilt of the tensor spectrum depends on certain assumptions and parameter choices, as we will describe in the next section.

%%%%%%%%%%%%%%%%%%%%%%%%%%%%%%%%%%%%%%%%%%%%%%%%%%%%%%%%%%%%%%%%%%%%%%%%%%%%%%%%%%%%%%%%%%%%%%%%%%%%
%%%%%%%%%%%%%%%%%%%%%%%%%%%%%%%%%%%%%%%%%%%%%%%%%%%%%%%%%%%%%%%%%%%%%%%%%%%%%%%%%%%%%%%%%%%%%%%%%%%%

\section{Models with blue tilt}\label{sec4}

In this section, we investigate in detail several quantum-gravity models predicting blue-tilted GW spectra at CMB scales. We predict the GW amplitude at interferometer scales for each model and compare the theoretical prediction of the GW spectrum with the sensitivity curves of LIGO-Virgo-KAGRA (LVK) \cite{TheLIGOScientific:2016wyq,Abbott:2017xzg,Akutsu:2018axf}, LISA \cite{Bartolo:2016ami,Caprini:2019pxz}, Einstein Telescope (ET) \cite{Maggiore:2019uih}, DECIGO \cite{Seto:2001qf,Kawamura:2011zz,Kawamura:2020pcg} and the future pulsar timing array project by SKA \cite{Janssen:2014dka}.

%%%%%%%%%%%%%%%%%%%%%%%%%%%%%%%%%%%%%%%%%%%%%%%%%%%%%%%%%%%%%%%%%%%%%%%%%%%%%%%%%%%%%%%%%%%%%%%%%%%%

\subsection{Non-local Starobinsky inflation}\label{nostar}

We briefly recall that one of the most favored inflationary models to date \cite{Akrami:2018odb} is local Starobinsky inflation, based on the Lagrangian \cite{Starobinsky:1980te,Mukhanov:1981xt,Barrow:1983rx,Starobinsky:1983zz,Whitt:1984pd,Kofman:1985aw,Starobinsky:1987zz,Maeda:1987xf,Maeda:1988rb}
\be\label{starol}
\cL=\frac{\Mpl^2}{2}\left(R+\frac{R^2}{6m^2}\right),
\ee
where $m$ is a mass scale. While this cosmological model is not based on a local theory of quantum gravity because, contrary to local Stelle gravity \cite{Ste77,Ste78,Julve:1978xn,Fradkin:1981iu,Fradkin:1981hx,Avramidi:1985ki,HOW}, it does not contain Riemann-tensor and Ricci-tensor terms, it can be embedded in non-local quantum gravity. Non-local quantum gravity is a covariant, perturbative quantum field theory of gravity which is unitary and super-renormalizable or finite \cite{Kra87,Kuz89,Tom97,Modesto:2011kw,BGKM,BCKM,MoRa1,MoRa2,Buoninfante:2018mre} (see \cite{Modesto:2017sdr} for a review). The fundamental action is second-order in curvature tensors and infinite-order in spacetime derivatives. 
%The classical action of non-local quantum gravity in the Weyl basis in $D=4$ topological dimensions is
 As is well known from the classic example of local Stelle gravity, having second-order curvature terms makes the theory renormalizable (i.e., ultraviolet divergences are under control, there is only a finite number of divergent Feynman diagrams, and the theory is predictive because it contains a finite number of coupling constants). However, the four derivatives included in the second-order curvature terms are responsible for unstable modes and the quantum theory is not unitary. Probability is conserved, and renormalizability is improved, when suitable nonlocal operators called form factors are inserted in the definition of the classical Lagrangian, which is
\be\label{TWeyl}
S =\frac{\Mpl^2}{2}\int\rmd^4x\,\sqrt{|g|}\left[R+R\,\g_{\rm S}(\B)R+C_{\mu\nu\rho\s}\g_{\rm C}(\B)C^{\mu\nu\rho\s}+V(C)\right],
\ee
where $C_{\mu\nu\rho\s}$ is the Weyl tensor and $V(C)$ is a set of local terms in the Weyl tensor necessary if one wants to make the quantum theory finite instead of just renormalizable. They play no role in the cosmological model because $C$ is identically zero on a FLRW background and the operators in $V$ do not enter the graviton propagator. The form factors $\g_{\rm S}$ and $\g_{\rm C}$ are chosen in such a way that the theory be renormalizable and free of ghosts on a given background. For instance, on Minkowski spacetime
\be
\g_{\rm S}(z) =-\frac{1}{6M_*^2z}\left[\rme^{\H_0(z)}\left(1-\frac{M_*^2}{m^2}z\right)-1\right],\qquad \g_{\rm C}(z) = \frac{1}{2M_*^2}\frac{\rme^{\H_2(z)}-1}{z}\,,\label{gSC}
%\g_{\rm S}(\B) =-\frac{1}{6\B}\left[\rme^{\H_0(\B)}\left(1-\frac{\B}{m^2}\right)-1\right],\qquad \g_{\rm C} = \frac{\rme^{\H_2(\B)}-1}{2\B}\,,
\ee
where $\H_{0,2}(z)$ are entire functions of the dimensionless combination
\be
z:=\frac{\B}{M_*^2}\,,
\ee
and $M_*$ is the fundamental mass of the theory. A very general class of functions $\H_{0,2}$ with benign UV properties are asymptotically polynomial, i.e., they vanish at small $z$ and go as the logarithm of a polynomial of degree $n_{\rm deg}$ at large $z$:
\be\label{asimp}
\lim_{z\to 0} \H_{0,2}(z)=0\,,\qquad \lim_{z\to\infty} \H_{0,2}(z)=\ln |z|^{n_{\rm deg}}\,,
\ee
where $n_{\rm deg}\geq 1$. Therefore, at high energies, $\exp\H_{0,2}\sim |z|^{n_{\rm deg}}$. The mass $m$ in \Eq{gSC} satisfies the hierarchy $m\ll M_*\lesssim\Mpl$ during inflation. It is associated with a scalar degree of freedom introduced to recover the Starobinsky Lagrangian \Eq{starol} in the local limit (expansion and truncation of the form factors to leading order \cite{Briscese:2013lna}) but the theory is UV finite and unitary also in the limit $m\to\infty$.

On curved backgrounds, the expressions for $\g_{\rm S}$ and $\g_{\rm C}$ change. In particular, on a quasi de Sitter background with Ricci scalar $\bar R\approx {\rm const}$, the absence of extra ghost degrees of freedom requires \cite{Koshelev:2016xqb,SravanKumar:2018dlo}
\ba
\g_{\rm S}(z) &=&-\frac{1}{6M_*^2(z+2z_*)}\left[\rme^{\H_0(z+2z_*)}\left(1-\frac{M_*^2}{m^2}z\right)-\left(1+\frac{2M_*^2}{m^2}z_*\right)\right]\,,\label{gS2}\\
\g_{\rm C}(z) &=&\left(\frac{z_*}{m^2}+\frac{1}{2M_*^2}\right)\frac{\rme^{\H_2(z-4z_*)}-1}{z-4z_*}\,,\label{gC2}
\ea
where 
\be\label{zst}
z_*:=\frac{R}{6M_*^2}\,.
\ee
Equations \Eq{gS2} and \Eq{gC2} are presented as valid for any curved metric but, strictly speaking, they guarantee ghost freedom only for de Sitter and there is no proof that they would work also for other backgrounds. For this reason, in the following one should regard \Eq{gS2} and \Eq{gC2} as  evaluated on the quasi de Sitter background ($\B=\bar\B$ in $z$ and $R=\bar R$ in $z_*$). In the limit $z_*\to 0$ (Minkowski spacetime), equations \Eq{gS2} and \Eq{gC2} reduce to \Eq{gSC}.

From \textsc{Planck} data, at the CMB pivot scale \Eq{pivot} one can write the mass $m$, the Hubble parameter at horizon crossing and the background Ricci scalar at horizon crossing in terms of the reduced Planck mass $\Mpl$ and the number of e-foldings corresponding to the CMB pivot scale $\cN_{\rm piv}$ \cite{Koshelev:2016xqb,Koshelev:2020foq}:
\be
m \simeq \frac{7.15\times 10^{-4}}{\cN_{\rm piv}}\,\Mpl\,,%\qquad H_* \simeq \frac{3.08\times 10^{-3}}{\cN}\,\Mpl\,,
\qquad
\bar R \simeq \frac{12.1\times 10^3}{\cN_{\rm piv}}\,m^2\,,\label{mRN}
\ee
so that
\be
z_*\simeq \frac{1.03\times 10^{-3}}{\cN_{\rm piv}^3}\frac{\Mpl^2}{M_*^2}\,.
\ee
For Starobinsky inflation, we find $\cN_{\rm piv} \approx 54$.

On a de Sitter or quasi de Sitter background, the point $z=z_*\neq 0$ corresponds to the pole of the graviton propagator read from the second-order Lagrangian%, which gives the equation of motion,
 for the transverse tensor metric perturbation $h_{\mu\nu}=g_{\mu\nu}-\bar g_{\mu\nu}$:
\be\label{S2}
\de^{(2)}S=\frac{\Mpl^2}{2}\int\rmd^4x\,\sqrt{|g|}\,h_{ij}\left(\bar\B-\frac{\bar R}{6}\right)\,\rme^{\tilde\H_2(\bar\B)}h^{ij}\,,
\ee
where\footnote{Our notation with respect to previous papers is $\tilde\H_2(z)=-2\om(z)$ \cite{Koshelev:2016xqb}, $\tilde\H_2(z)=2\om(z)$ \cite{Koshelev:2017tvv} and $\tilde\H_2(z)=2\g_T(z-4z_*)$ \cite{Koshelev:2020foq}, while \Eq{tildeH} connects with \cite{SravanKumar:2018dlo}. Also, notice in $\g_{\rm C}$ the constant prefactor $1/(2M_*^2)$, absent in \cite{Koshelev:2016xqb,Koshelev:2017tvv} but present in \cite{SravanKumar:2018dlo}.}
\be\label{tildeH}
\tilde\H_2(z):=\H_2(z-4z_*)\,.
\ee
Note that all the expressions in this section are in the Jordan frame, contrary to the more common Einstein-frame formulation of local Starobinsky gravity and inflationary models \cite{CQC,DeFelice:2010aj}. The physical observables must be frame invariant 
%and, in fact, when expressed in terms of the number of e-folds they agree, since $\cN\simeq \cN_{\rm E}$ during inflation, where E denotes the Einstein frame 
\cite{DeFelice:2010aj}.
The only difference with respect to the kinetic term of the graviton in the Jordan frame in standard Starobinsky gravity is the exponential $\exp\tilde\H_2$.

The slow-roll approximation holds, since the background accelerating solution of local or non-local Starobinsky inflation is (quasi) de Sitter \cite{Koshelev:2016xqb}:
\be\label{aN}
a\simeq a_{\rm c}(t_{\rm e}-t)^{-\frac16}\,\rme^{-\frac{m^2}{12}(t_{\rm e}-t)^2}\qquad\Longrightarrow\qquad \cN\simeq \frac{m^2}{12}(t_{\rm e}-t)^2,
\ee
where $t_{\rm e}$ is the time at the end of inflation, $a_{\rm c}$ is a constant of energy dimension $-1/6$, and this and the following expressions are valid when $m(t_{\rm e}-t)\gg 1$. Since 
\be\label{aN2}
H\simeq \frac{m^2}{6}(t_{\rm e}-t)+\frac{1}{6(t_{\rm e}-t)}\,,
\ee
$\dot H\simeq -m^2/6$ and $\ddot H\simeq 1/[3(t_{\rm e}-t)^3]$, from equations \Eq{epseta} and \Eq{aN} one has
\be
\e \simeq \frac{1}{2\cN}\,,\qquad \eta \simeq \frac{1}{24\cN^2}\ll\e\,.
\ee

For later use, we will recast the Hubble parameter at horizon crossing in terms of the comoving wave-number. From \Eq{aN}, $t_{\rm e}-t\simeq \sqrt{12\cN}/m$, so that \Eq{aN2} becomes $H\simeq m\sqrt{\cN/3}+m/(12\sqrt{3\cN})$. Since the Hubble parameter is approximately constant during inflation, from definition \Eq{defiN} one has $\cN\simeq\ln(k_{\rm e}/k)$, where $k_{\rm e}$ is the comoving wave-number of the last perturbation crossing the horizon at the end of inflation. Therefore,
\be\label{Hk}
H(k)\simeq\frac{m}{\sqrt{3}}\left[\left(\ln\frac{k_{\rm e}}{k}\right)^{\frac12}+\frac{1}{12}\left(\ln\frac{k_{\rm e}}{k}\right)^{-\frac12}\right].
\ee
%Notice that $m$ is a constant in \Eq{Hk} and one should not replace $\cN$ in \Eq{mRN} with $\ln(k_{\rm e}/k)$. In fact, \Eq{mRN} are expressions derived from observational bounds and $\cN$ therein is a constant.

The primordial tensor spectrum $\cP_{\rm t}$ and the tensor-to-scalar ratio are \cite{Koshelev:2016xqb,Koshelev:2017tvv}
\ba
\cP_{\rm t} &\simeq& \frac{m^2}{2\pi^2\Mpl^2}(1-3\e)\,\rme^{-\tilde\H_2(z_*)}\,,\label{Ptfull}\\
r &\simeq&\frac{12}{\cN^2}\,\rme^{-\tilde\H_2(z_*)}\,.\label{ttsr}
\ea
The expression for $\cP_{\rm t}$ is in the next-to-leading order in the slow-roll parameters because, for illustrative purposes, we want to separate the standard part (local Starobinsky inflation, next-to-leading order contribution) and the non-local part (leading order contribution).

To calculate the tensor spectral index $n_{\rm t}$ \cite{Koshelev:2017tvv} and its running $\a_{\rm t}$, we use the slow-roll formalism and convert the final expression in terms of $\cN$ at the end. 
%We keep the leading order slow-roll expressions in the standard and the non-local parts separately, the latter being characterized by $O(z_*)$ contributions. Therefore, the standard part of the tensor spectrum is first-order in the slow-roll parameters and zero-order in $z_*$, while the non-local part is zero order in the slow-roll parameters and first order in $z_*$. The tensor index and the index running successively increase both orders by 1 because, on one hand, they are defined through the wave-number derivative \Eq{tk} at horizon crossing and, on the other hand, the time derivative of $\e$ is second-order, equation \Eq{dote}. We will also use
Using
\be\label{zsH}
z_*\simeq 2\left(\frac{H}{M_*}\right)^2\,,\qquad \frac{1}{H}\dot z_* \simeq  -2\e z_*\simeq -\frac{z_*}{\cN}\,,
\ee
we obtain the tensor spectral index \footnote{We take into account the fact that $z_*$ is a background-dependent quantity, which means that it is a constant on de Sitter but a varying function at next-to-leading order in the slow-roll approximation. Therefore, differentiation is performed directly on $\tilde\H_2(z_*)=\H_2(-3z_*)$, consistently with the slow-roll formalism. Since $\p_{z_*}\H_2(-3z_*)\neq \p_z\H_2(z-4z_*)|_{z=z_*}$, our expression for $n_{\rm t}$ differs with respect to \cite{Koshelev:2017tvv}, where $z_*$ in $\H_2(z-4z_*)$ is treated as a fixed number.}
\ba
n_{\rm t} &\simeq& -3\frac{\dot \e}{H}-\tilde\H_2'(z_*)\frac{\dot z_*}{H}\simeq -6\e^2+2\e z_* \tilde\H_2'(z_*)\nonumber\\
&\simeq& -\frac{3}{2\cN^2}+\frac{1}{\cN}z_* \tilde\H_2'(z_*)\,,\label{ent}
\ea
where a prime denotes the first derivative with respect to $z_*$. Similarly, the running of the tensor index is
\ba
\a_{\rm t} &\simeq& \frac{2}{H}[-6\e\dot\e+(\dot\e z_*+\e\dot z_*)\tilde\H_2'(z_*)+\e z_* \dot z_*\tilde\H_2''(z_*)]\nonumber\\
&\simeq& -4[6\e^3+\e\eta z_*\tilde\H_2'(z_*)+\e^2 z_*^2\tilde\H_2''(z_*)]\nonumber\\
&\simeq& -\frac{3}{\cN^3}-\frac{1}{12\cN^3} z_*\tilde\H_2'(z_*)-\frac{1}{\cN^2} z_*^2\tilde\H_2''(z_*)\,.\label{run}
\ea
Note that the last term dominates in the slow-roll expansion and is higher-order in $z_*\ll 1$. 
More generally, the $l$-th-order observable in the slow-roll approximation is proportional to the $l$-th derivative of the form factor,
\be\label{obs}
\cO_l:=\frac{\rmd^l\ln\cP_{\rm t}}{(\rmd\ln k)^l}\simeq -\left(-\frac{z_*}{\cN}\right)^l \tilde\H_2^{(l)}(z_*)+O\left(\frac{1}{\cN^{l+1}}\right),
\ee
where $\cO_0=\ln\cP_{\rm t}$, $\cO_1=n_{\rm t}$, $\cO_2=\a_{\rm t}$, and so on. The second term, sub-dominant, contains the standard slow-roll terms plus some new terms of the same order in $\cN$.

As we will see in the following, depending on the choice of the form factor $\H_{2}(z_*)$, we can obtain positive values of $n_t$ and $\a_t$ at the CMB scale, which might indicate a large GW amplitude at interferometer scales. However, this is not the case. We will study the GW spectrum using both the power-series approximation \Eq{Ptk} and the full expression \Eq{ttsr} and show that, in fact, the former does not give correct predictions at high frequencies.

\subsubsection{Kuz'min and Tomboulis form factors}

Let us now discuss the form factors $\H_{0,2}(\B)$. They must obey the conditions $\H_{0,2}(0)=0=\H_{0,2}(m^2)$. Since $\H_0$ plays no role in \Eq{S2}, we will not consider it in detail. The form of $\tilde\H_2$ is determined by (i) the allowed form factors in the full theory on Minkowski spacetime with $m\to\infty$, (ii) their generalization to a finite $m$ on Minkowski spacetime, (iii) their generalization to a finite $m$ and a curved background.

On Minkowski spacetime, the quantum theory is especially well defined in the UV in the case of two weakly non-local operators of the asymptotically polynomial type, one proposed by Kuz'min \cite{Kuz89}
\be
\H_{\rm Kuz}(z):=\a\left\{\ln p(z)+\Gamma[0,p(z)]+\gamma_\textsc{e}\right\},\label{kuz}
\ee
and the other by Tomboulis \cite{Tom97,Modesto:2011kw},
\be
\H_{\rm Tom}(z):=\frac{1}{2}\left\{\ln p^2(z)+\Gamma[0,p^2(z)]+\gamma_\textsc{e}\right\},\label{tom}
\ee
where $\a\geq 3$, $\Gamma$ is the upper incomplete gamma function with its first argument vanishing, $\gamma_\textsc{e}\approx 0.577$ is the Euler--Mascheroni constant and $p(z)$ is a polynomial such that ${\rm Re}\,p(z)>0$.

The generalization of \Eq{kuz} and \Eq{tom} to a Starobinsky term is non-trivial and it involves a choice of polynomial $p(z)$ such that UV divergences in momentum integrals are under control (renormalizability) or disappear altogether (finiteness). For instance, in the case of Tomboulis form factor the beta functions of Newton's constant and the cosmological constant are identically zero and the theory is super-renormalizable provided $p(\B)\propto m^2\B^{q-5}(\B^3-m^6)^2/(\B-m^2)$ for $q\geq 6$ \cite{Koshelev:2016xqb}. The proportionality coefficient $m^2$ is inserted in order to cancel the coefficient in the full form factor and thus ensuring that the beta function of $m$ is a constant \cite{Koshelev:2016xqb}.\footnote{A trivial running of $m$ may help to preserve asymptotic freedom and accessibility to analytic calculations in the quantum theory, but this condition could be relaxed and we do not delve into it here.} 

Things become more complicated on curved backgrounds not only because the translation $p(z)\to p(z-4z_*)$ should be combined with appropriate $m$ factors, but also because the presence of the Ricci scalar in $z_*$ renders the theory non-renormalizable. The first issue is not particularly pressing. Although the $m$-dependence has not been worked out on a curved background, one can ignore it because during inflation $m$ is much smaller than the other scales. The second issue is more serious for the field-theoretician, albeit not necessarily for the cosmologist. In flat space, a derivative expansion of $C\g_{\rm C} C$ yields an $O(\B^{2+n})$ leading term, where $n$ is an integer. This is the same order in the propagator $\sim 1/(k^2)^{2+n}$ and they compensate each other in Feynman diagrams. However, when $R$ factors are present in $\g_{\rm C}$ the derivative order increases by 2 and interactions become $O(4+n)$, thus carrying a divergence worse than the propagator. Until superseded by a more formal proof, this heuristic argument suggests that this cosmological model is not fully embedded in a theory of quantum gravity on curved spacetimes.

\begin{figure}
\centering
\includegraphics[scale=0.9]{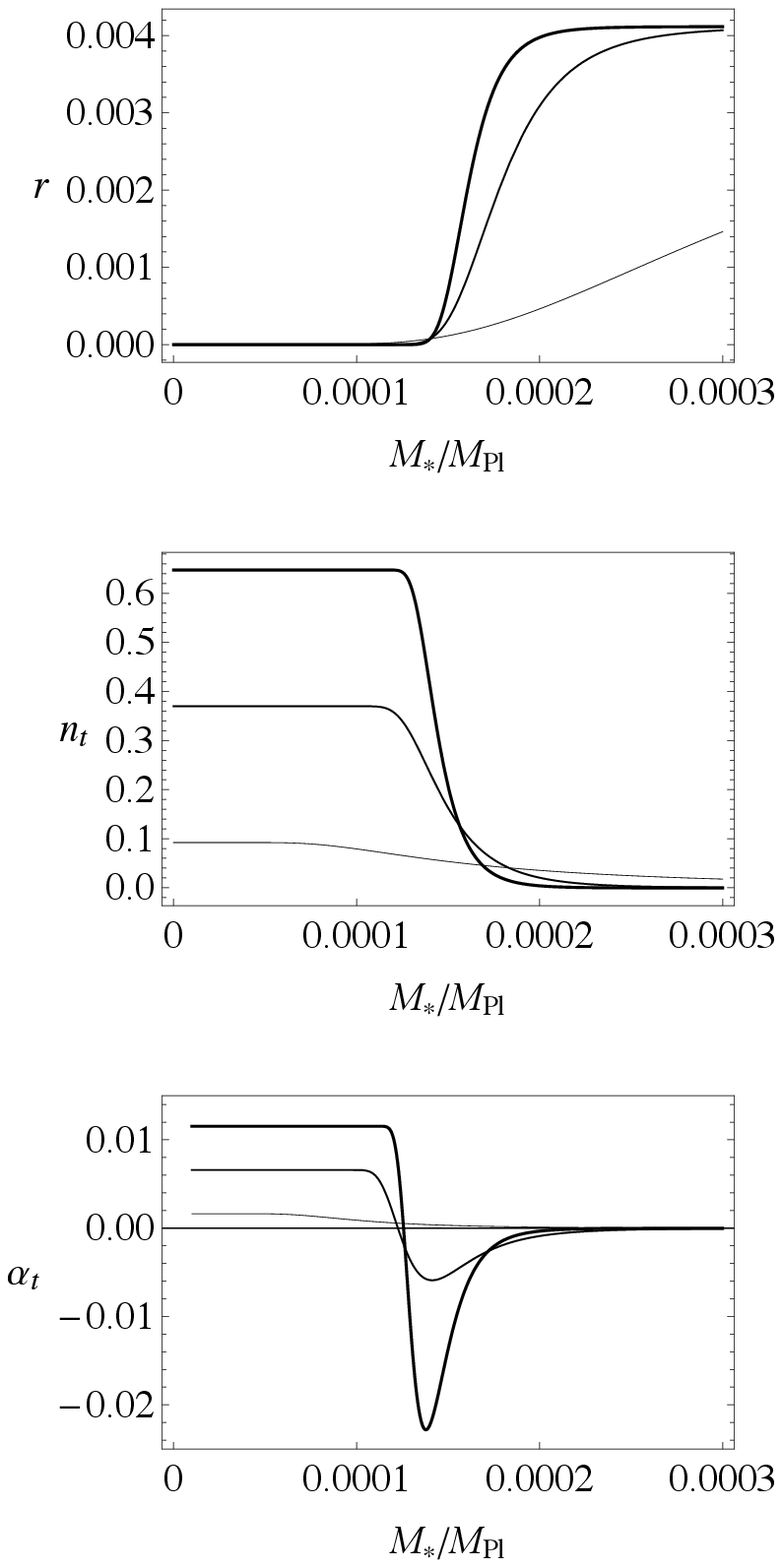}\includegraphics[scale=0.9]{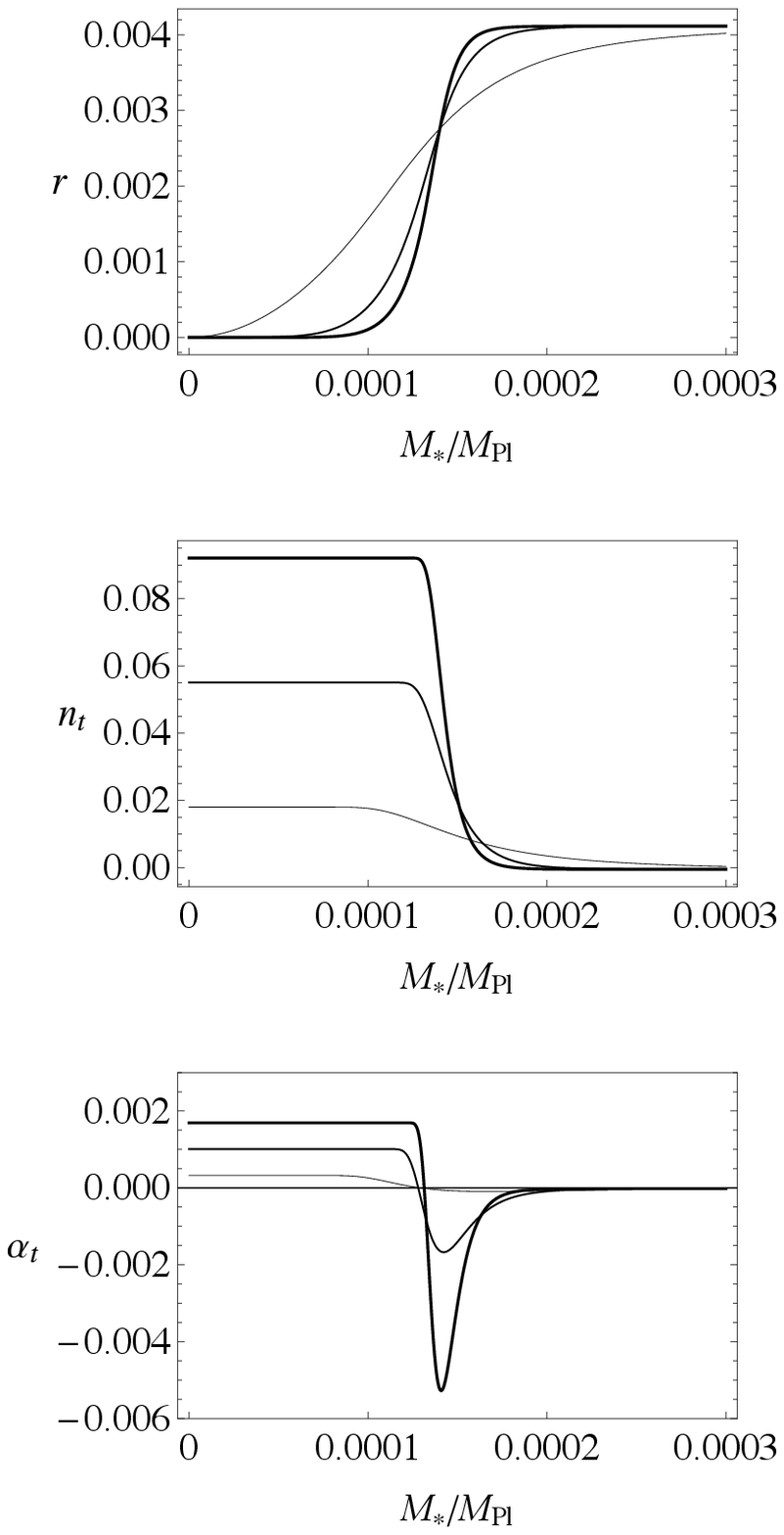}
\caption{\label{fig1} Tensor-to-scalar ratio $r$ \Eq{ttsr}, tensor spectral index $n_{\rm t}$ \Eq{ent} and tensor running $\a_{\rm t}$ \Eq{run} in non-local Starobinsky inflation with Kuz'min form factor \Eq{kuz} (left plots) or Tomboulis form factor \Eq{tom} (right plots) with a polynomial $p(z)=(-z)^q$ as a function of the ratio $M_*/\Mpl$, for $\cN_{\rm piv}=54$. Left plots: $\a=5$ and $q=1,4,7$ (increasing thickness). Right plots: $q=1,3,5$ (increasing thickness).}
\end{figure}

With this caveat on board, we can still extract phenomenology from the form factors of the theory. The tensor-to-scalar ratio $r$,  the tensor index $n_{\rm t}$, and the tensor running $\a_{\rm t}$ stemming from $\H_{\rm Kuz}(z-4z_*)$ and $\H_{\rm Tom}(z-4z_*)$ with $p(z)=(-z)^q$ are shown in Fig.\ \ref{fig1}.
%In both cases, for $M_*/\Mpl<10^{-4}$ the tensor index saturates to a finite value. Kuz'min form factor can generate a higher tensor index more easily than Tomboulis', but both have an extremely low tensor-to-scalar ratio in the range where $n_{\rm t}$ is maximal. One can show that the running of $n_{\rm t}$ is also positive but low in that region and it cannot boost the tensor index much. These results are reported only for the sake of establishing a contrast between the conclusions one could naively reach by examining slow-roll observables and the correct ones we will get from the full tensor spectrum.
Although the tensor-to-scalar ratio tends to be very low when $n_{\rm t}$ and $\a_{\rm t}$ are maximal, there is a corner of the parameter space where we can expect large values of $n_{\rm t}$ and $\a_{\rm t}$ which could enhance the GW amplitude at interferometer scales. However, as we will see below, we should be careful about using the power-series parametrization \Eq{Ptk} by $n_{\rm t}$ and $\a_{\rm t}$.

%The case of asymptotically polynomial form factors is more complicated. We can get some insight noting that $z_*<1$. 
Equation \Eq{obs} signals a possibly dangerous problem of convergence of the parametrization \Eq{Ptk} and its higher-order generalizations. Whenever the form factor is such that $\tilde\H_2^{(n)}(z_*)$ is large, one cannot use the tensor index and the running as a reliable description of the spectrum. Let us condense Kuz'min and Tomboulis form factors \Eq{kuz} and \Eq{tom} into a single expression $\H_{\rm pol}(z)=\a\left\{\ln p^s(z)+\Gamma[0,p^s(z)]+\gamma_\textsc{e}\right\}$, where $\a=1/2$ and $s=2$ for Tomboulis while $\a\geq 3$ and $s=1$ for Kuz'min. Expanding for small $z$ (small $p$), one has $\H_{\rm pol}(z)=\a\sum_{j=1}^{+\infty} (-p^s)^j/(j!j)$, so that for $p(z)=(-z)^q$ one has
\be
\tilde\H_{\rm pol}^{(l)}(z_*)=\a \sum_{j=1}^{+\infty} (-1)^j3^{qsj} \frac{qsj(qsj-1)\cdots (qsj-l+1)}{j!j} z_*^{qsj-l}\,.
\ee
The leading-order term in $\cO_l$ is $\propto[\a(-1)^j3^{l}l!/(j!j)] z_*^l$, where $j=l/(qs)$. This term will be sizable for an arbitrary order $l$ unless $z_*\ll 1$, a regime where the non-local effect is small. Therefore, for this class of form factors it is not possible to use the parametrization \Eq{Ptk} or any of its generalizations, and one must rely on the exact expression of $\cP_{\rm t}(k)$. The latter is given by \Eq{Ptfull} with $z_*(k)$ given by \Eq{zsH} and \Eq{Hk}:
\be
\cP_{\rm t}(k) \simeq \frac{m^2}{2\pi^2\Mpl^2}\left[1-\frac{3}{2\ln(k_{\rm e}/k)}\right]\,\rme^{-\tilde\H_2\left[\frac{2H^2(k)}{M_*^2}\right]}\,.\label{Ptfull2}
\ee

\begin{figure}
\centering
\includegraphics[width=10cm]{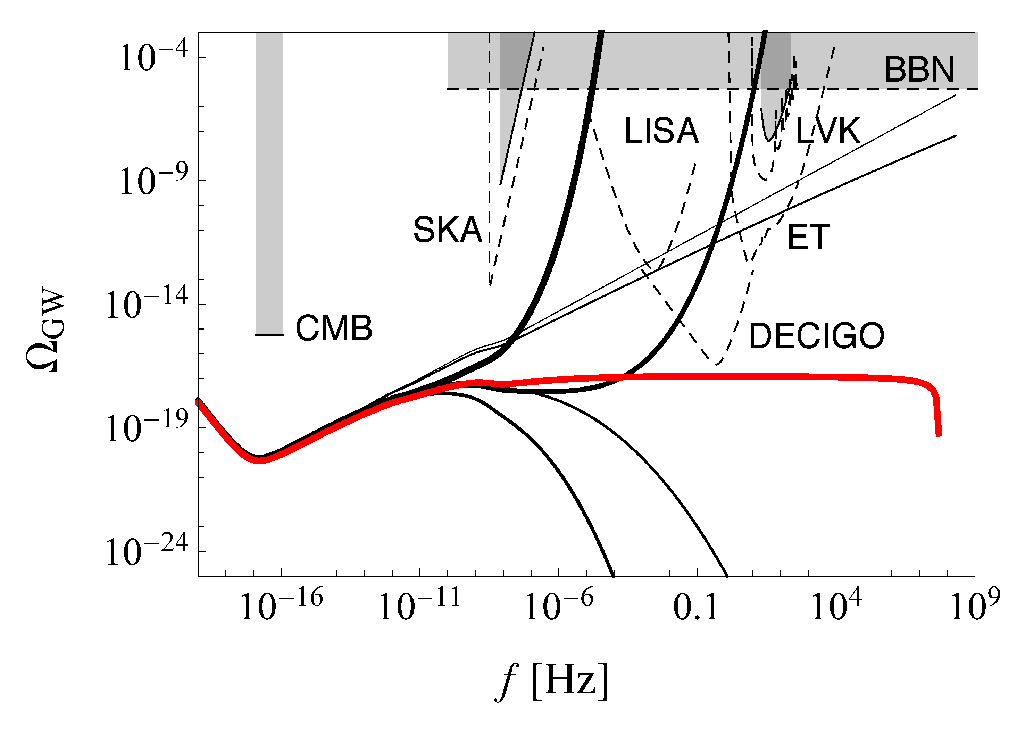}
\caption{\label{fig2} 
SGWB of non-local Starobinsky inflation with Kuz'min form factor \Eq{kuz} with $q=7$ and $\a=5$. We compare the GW spectra of the power-series parametrization \Eq{Ptk} (black solid curves) with the full expression \Eq{Ptfull} (red solid curve). The slow-roll approximation of the 1st to 6th order are calculated using \Eq{obs}; the thickness of the line increases with the order. We also show the sensitivity curves of LIGO-Virgo-KAGRA (LVK), SKA, LISA, ET and DECIGO. Shaded regions are excluded by CMB ($r<0.068$), pulsar timing (NANOGrav 11-year data \cite{Arzoumanian:2018saf}), gravitational-wave (current LIGO data) and BBN observations. Note that we widened the single-frequency CMB constraint to a band in order to make it more visible.}
\end{figure}

Using Kuz'min form factor with $q=7$ and $\a=5$ as an example, we compare the power-series parametrization \Eq{Ptk} with the full expression of the spectrum \Eq{Ptfull2} in Fig.~\ref{fig2}. As we can see, the power-series expression by $n_{\rm t}$ and $\a_{\rm t}$ predicts a very large amplitude which could reach the sensitivity of several interferometer experiments. However, as expected from the above discussion, we see that the spectrum does not converge by adding the higher-order terms \Eq{obs}, contrary to the case of the standard inflationary scenario \cite{Kuroyanagi:2011iw}. Comparing with the full spectrum, we find that the parametrization \Eq{Ptk} is good near the CMB scale but deviates largely from the true spectrum at high frequencies. 

\begin{figure}
\centering
\includegraphics[width=7.8cm]{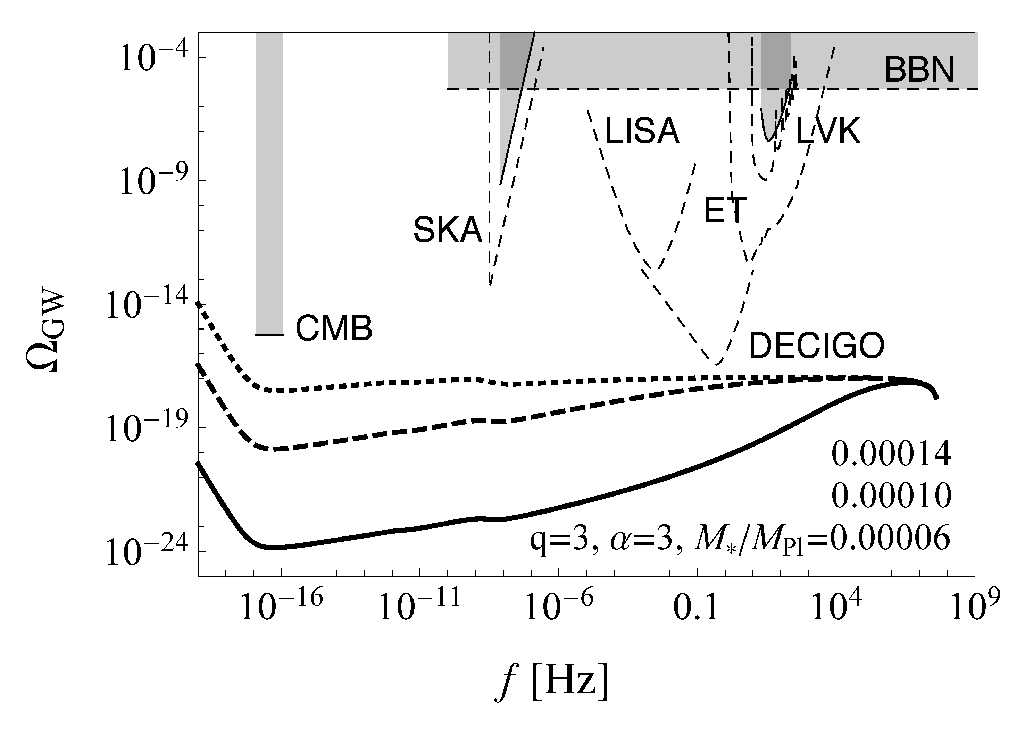}
\includegraphics[width=7.8cm]{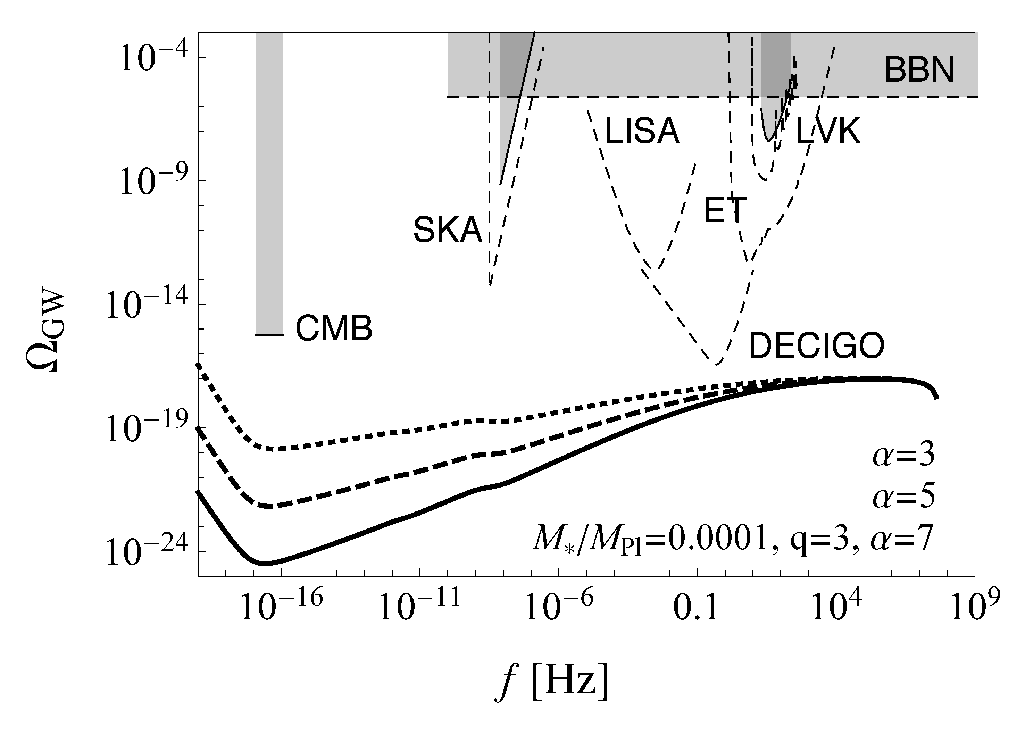}\includegraphics[width=7.8cm]{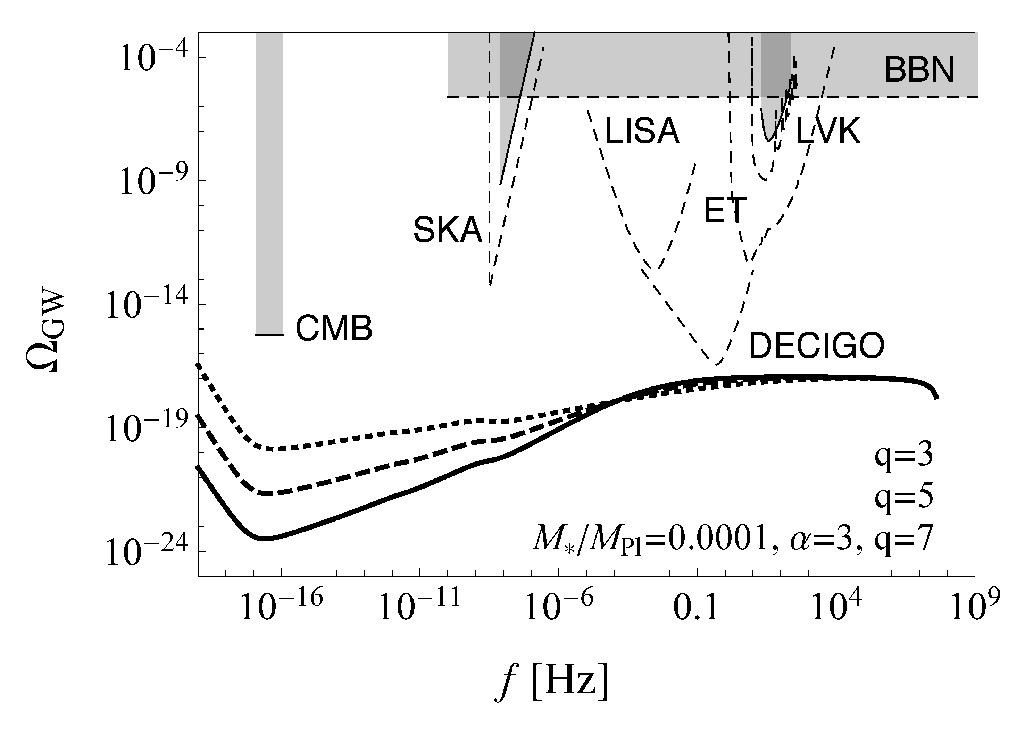}
\caption{\label{fig3} SGWB of non-local Starobinsky inflation with Kuz'min form factor \Eq{kuz} for different values of $M_*$, $q$ and $\a$. Top panel: $q=3$, $\a=3$ and $M_*/\Mpl=0.00014, 0.0001, 0.00006$ (respectively, dotted, dashed and solid black curve). Bottom left panel: $M_*/\Mpl=0.0001$, $q=3$ and $\a=3,5,7$ (respectively, dotted, dashed and solid black curve). Bottom right panel: $M_*/\Mpl=0.0001$, $\a=3$ and $q=3,5,7$ (respectively, dotted, dashed and solid black curve).}
\end{figure}

\begin{figure}
\centering
\includegraphics[width=7.8cm]{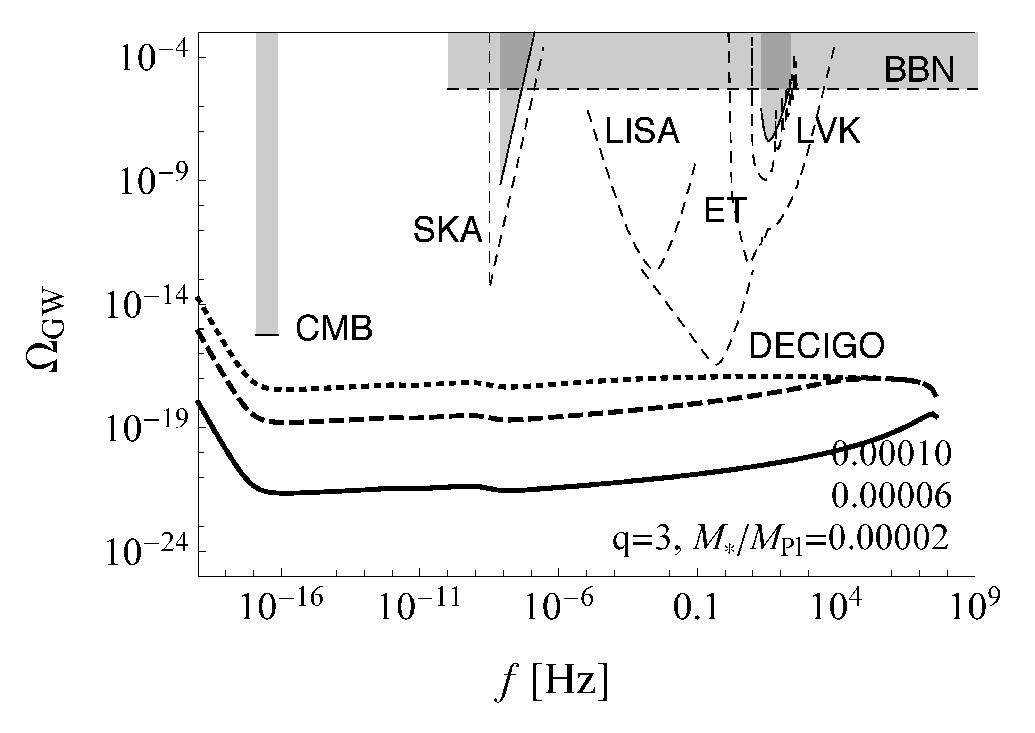}\includegraphics[width=7.8cm]{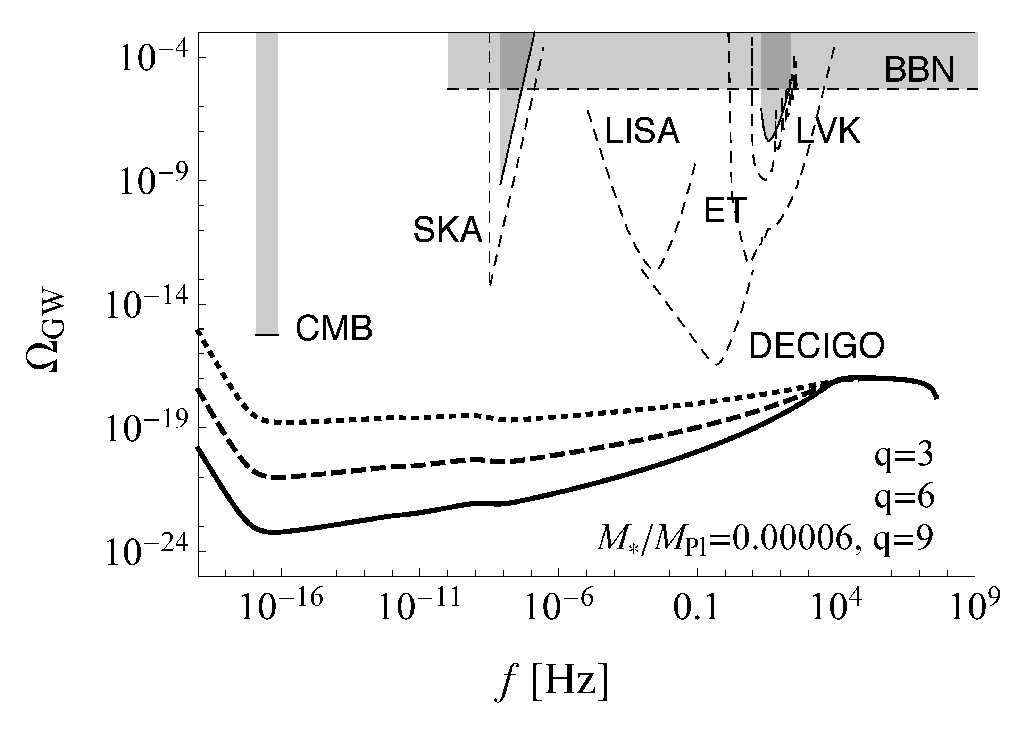}
\caption{\label{fig4} SGWB of non-local Starobinsky inflation with Tomboulis form factor \Eq{tom} for different values of $M_*$ and $q$, compared with the sensitivity curves of GW experiments. Left panel: $q=3$, and $M_*/\Mpl=0.0001, 0.00006, 0.00002$  (respectively, dotted, dashed and solid black curve). Right panel: $M_*/\Mpl=0.00006$, and $q=3,6,9$ (respectively, dotted, dashed and solid black curve).}
\end{figure}

The important message here is that \emph{a blue spectrum at the CMB scale does not always mean observable GWs at interferometer scales}. In Figs.~\ref{fig3} and \ref{fig4}, we show the SGWB spectrum of, respectively, Kuz'min and Tomboulis form factors for different parameter values, calculated using the full expression \Eq{Ptfull2}. Figure \ref{fig1} suggests that we should consider a small ratio $M_*/\Mpl$ in order to maximize the chance of a blue tilt. On the other hand, recalling the discussion below \Eq{asimp}, the lower bound of $M_*$ is $m$, which is about $10^{-5}\Mpl$ if we take \Eq{mRN} into account. Taking values of $M_*/\Mpl$ close to this lower bound, we find that the GW spectra are flattened at high frequencies in all cases, and there are no experiments having enough sensitivity to detect the GWs.
%We checked that the result is the same if one uses the numerical solution of the background equations of motion instead of the approximated expression \Eq{Hk}, in agreement with the comparison made in \cite{SravanKumar:2018dlo} at the level of the quasi-de Sitter solution. As one can see, the SGWB blue tilt with Tomboulis form factor is less pronounced than in Kuz'min case but the amplitude at CMB scales is larger.

%Both for asymptotically polynomial and for monomial form factors, 
The SGWB at high frequencies matches the ordinary Starobinsky case, i.e., a red-tilted undetectable spectrum. This result is highly non-trivial because a cursory study of the tensor index \Eq{ent}, with or without running \Eq{run}, would have led to a radically different conclusion. We can understand the intuitive reason why non-local corrections to the tensor spectrum \Eq{Ptfull2} vanish at high frequencies. Equation \Eq{Hk} tells us that the Hubble parameter is approximately proportional to a positive power of $\ln(k_{\rm e}/k)>0$ during inflation. Thus, both $H$ and $z_*\propto H^2$ decrease when $k$ increases. Since $\tilde\H_2$ increases with $z_*$, it decreases when $k$ increases and vanishes asymptotically. Therefore, at high frequencies (high $k$) the non-local term $\exp[-\tilde\H_2(z_*)]$ in the tensor spectrum \Eq{Ptfull} tends to unity,
\be
\lim_{k\to\infty} \rme^{-\tilde\H_2(z_*)}=1\,,
\ee
and $\cP_{\rm t}$ reduces to the standard Starobinsky spectrum at the frequencies of present and future GW interferometers.

%%%%%%%%%%%%%%%%%%%%%%%%%%%%%%%%%%%%%%%%%%%%%%%%%%%%%%%%%%%%%%%%%%%%%%%%%%%%%%%%%%%%%%%%%%%%%%%%%%%%

\subsubsection{Monomial form factor}
\begin{figure}
\centering
\includegraphics[width=7.8cm]{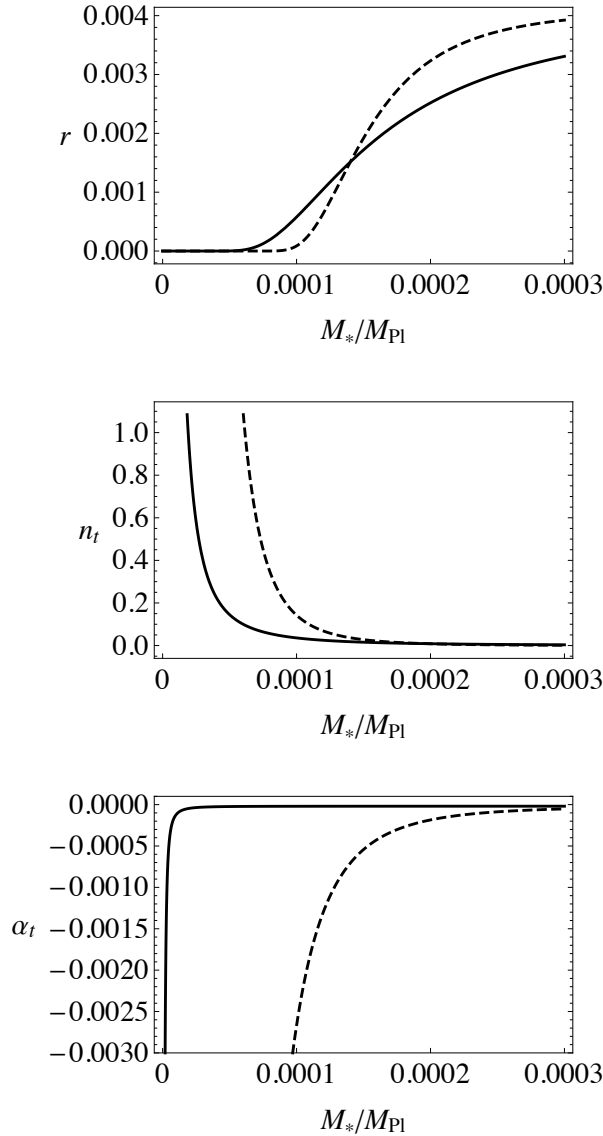}
\caption{\label{fig5} Tensor-to-scalar ratio \Eq{ttsr}, tensor spectral index \Eq{ent} and tensor running \Eq{run} with monomial form factor \Eq{mono} as a function of the ratio $M_*/\Mpl$, for $\cN=54$ and with $n=1$ (Wataghin form factor, solid curve) or $n=2$ (Krasnikov form factor, dashed curve).}
\end{figure}

Other form factors in non-local quantum gravity predict an unbounded tensor index. These are the monomials
\be\label{mono}
\H_{\rm mon}(z):=(-z)^n\,.
\ee
The case $n=1$ corresponds to Wataghin form factor \cite{Wataghin:1934ann}, typical in string field theory \cite{ohm01} but very often used also in non-local quantum gravity \cite{Biswas:2005qr,CaMo2}, while for $n=2$ one has Krasnikov form factor \cite{Kra87}. They both generate a strongly suppressed but also strongly blue-tilted tensor spectrum for $M_*/\Mpl<10^{-4}$, with negative running as shown in Fig.\ \ref{fig5}. %(Fig.\ \ref{fig3}). 
%We cannot determine from Fig.\ \ref{fig2} how much the negative running flattens the spectrum, an issue we will solve in section \ref{sec4}.

\begin{figure}
\centering
\includegraphics[width=7.8cm]{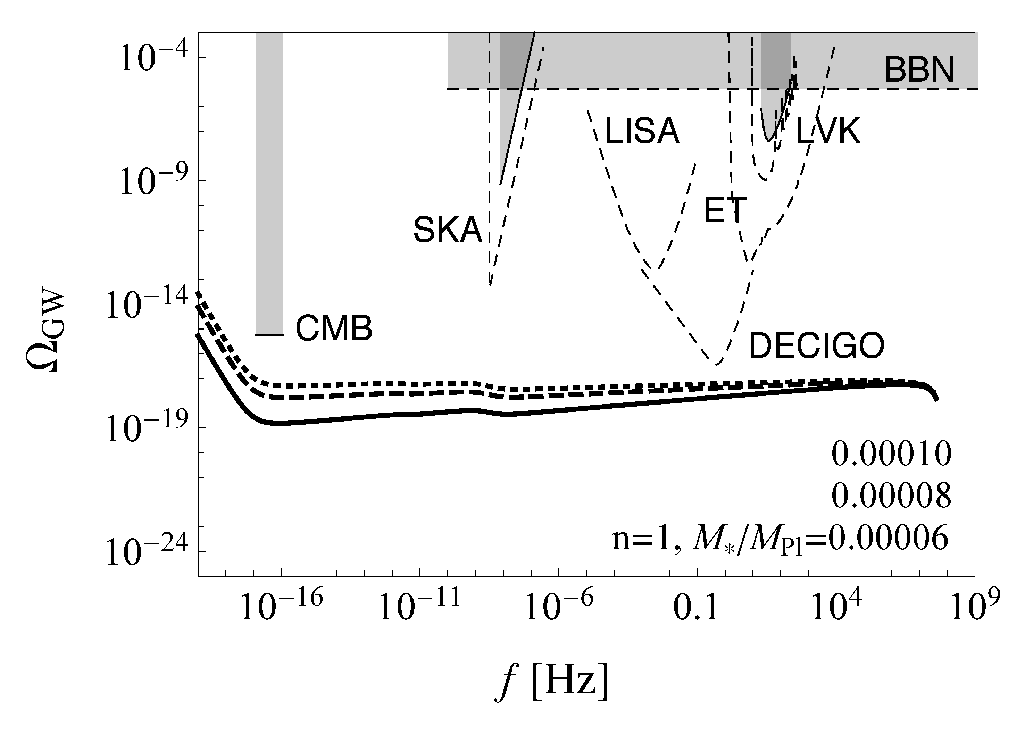}\includegraphics[width=7.8cm]{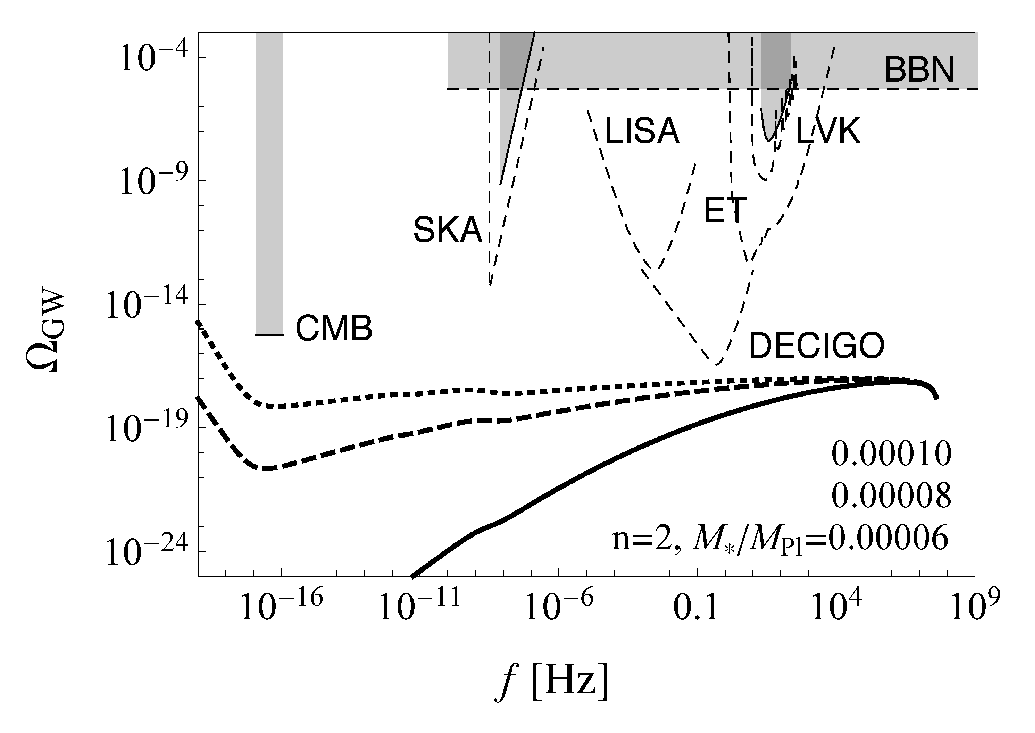}
\caption{\label{fig6} SGWB of non-local Starobinsky inflation with monomial form factor \Eq{mono} for $n=1,2$ (respectively, left and right panel) and $M_*/\Mpl=0.0001, 0.00008, 0.00006$  (respectively, dotted, dashed and solid black curve), compared with the sensitivity curves of GW experiments.}
\end{figure}

In Fig.\ \ref{fig6}, we plot the SGWB spectrum of the model with monomial form factor for different values of $M_*/\Mpl$, calculated using the full expression \Eq{Ptfull}. For the monomial form factors, the $l$-th derivative is given by $\tilde\H_{\rm mon}^{(l)}(z_*)=3^n n(n-1)\cdots(n-l+1)z_*^{n-l}$ and one should take into account all the first $l=n$ slow-roll observables. This means that the tensor index is sufficient to describe the spectrum with Wataghin form factor ($n=1$), while for Krasnikov form factor ($n=2$) one must also consider the tensor running and can safely ignore the running of the running. In fact, we find that the power-series parametrization is in good agreement with the full spectrum when we take into account up to $l=n$ terms. However, the behaviour of the spectrum at high frequencies is similar to the previous cases of Kuz'min and Tomboulis form factors and the SGWB never reaches the detectability threshold of any experiment. This can be understood in terms of the power-series spectrum that, even having a large blue tensor tilt, never has a large GW amplitude because the tensor-to-scalar ratio is highly suppressed at the CMB scale and, furthermore, is suppressed by the tensor running $\a_{\rm t}$ at high frequencies.
%where one can see that the tensor running $\a_{\rm t}$ (now a meaningful observable, as we argued at the end of section \ref{nostar}) causes a suppression of the spectrum. As a consequence, the SGWB never reaches the detectability threshold established by the DECIGO sensitivity curve. Changing the energy ratio $M_*/\Mpl$ can affect the tilt and amplitude considerably but never to the point of hitting the DECIGO range.

%%%%%%%%%%%%%%%%%%%%%%%%%%%%%%%%%%%%%%%%%%%%%%%%%%%%%%%%%%%%%%%%%%%%%%%%%%%%%%%%%%%%%%%%%%%%%%%%%%%%

\subsubsection{Other form factors}

One can concoct other phenomenological form factors \cite{SravanKumar:2018dlo,Koshelev:2020foq} and here we present two more that respect the asymptotic limits of weakly non-local operators: the difference between two Tomboulis form factors,
\be\label{tomtom}
\H_\textrm{Tom-Tom}:=\H_{\rm Tom,1}(z)-\H_{\rm Tom,2}(z)\,,\qquad {\rm deg}\, p_1(z)>{\rm deg}\, p_2(z)\,,
\ee
similar to what proposed in \cite{SravanKumar:2018dlo}, and the difference between Tomboulis and Kuz'min form factors,
\be\label{tomkuz}
\H_\textrm{Tom-Kuz}:=\H_{\rm Tom}(z)-\H_{\rm Kuz}(z)\,,\qquad {\rm deg}\, p^2_{\rm Tom}(z)>{\rm deg}\, p^a_{\rm Kuz}(z)\,.
\ee
Looking only at leading-order slow-roll observables, they both yield a moderately blue-tilted spectrum (same order as for Tomboulis form factor) with relatively high $r$ (much higher than Tomboulis). However, this blue tilt is an artifact just like for Kuz'min and Tomboulis form factors. %, as one can see in Fig.\ \ref{fig3}. Before reaching the plateau, $n_{\rm t}$ achieves a global maximum where $r$ is non-negligible.
% Since we will get interesting phenomenology already with the more common form factors \Eq{kuz}, \Eq{tom} and \Eq{mono}, we will not compare the theory with \Eq{tomtom} or \Eq{tomkuz} with experiments.
%\begin{figure}
%\centering
%\includegraphics[width=7.8cm]{fig3a}\includegraphics[width=7.8cm]{fig3d}\\
%\includegraphics[width=7.8cm]{fig3b}\includegraphics[width=7.8cm]{fig3e}%\\
%%\includegraphics[width=7.8cm]{fig3c}\includegraphics[width=7.8cm]{fig3f}
%\caption{\label{fig3} Tensor-to-scalar ratio \Eq{ttsr} and tensor spectral index \Eq{ent} %and tensor running \Eq{run}
% with form factor \Eq{tomtom} (left plots) or \Eq{tomkuz} (right plots) with $p(z)=(-z)^q$ as a function of the ratio $M_*/\Mpl$, for $\cN=55$. Left plots: $q_2=1$ and $q_1=2,3,5$ (increasing thickness). Right plots: $\a=3$, $q_{\rm Kuz}=1$ and $q_{\rm Tom}=2,3,5$ (increasing thickness).}
%\end{figure}

%A final remark about the figures is that the energy scales therein do not entail any fine tuning of $M_*$. The only constraint coming from particle physics is that $M_*>O({\rm TeV})$ \cite{Biswas:2014yia}, well below the range required for a detectable SGWB.

%%%%%%%%%%%%%%%%%%%%%%%%%%%%%%%%%%%%%%%%%%%%%%%%%%%%%%%%%%%%%%%%%%%%%%%%%%%%%%%%%%%%%%%%%%%%%%%%%%%%

\subsection{String-gas cosmology}\label{stgassec}

In this scenario, inflation is replaced by a quasi-static era \cite{Bernardo:2020nol,Bernardo:2020bpa} where thermal fluctuations generate an almost scale-invariant primordial scalar and tensor spectrum, whose derivation is reviewed in \cite{Brandenberger:2008nx,Bra11,Bra12,Brandenberger:2015kga}. The tensor spectrum is \cite{BrNPV,BNPV,BrNP}
\be\label{sgc2}
\cP_{\rm t}(k)\simeq\frac{1}{4(\Mpl l_{\rm st})^4}\hat T(k)\left[1-\hat T(k)\right]\,\ln^2\left[\frac{1-\hat T(k)}{l_{\rm st}^2k^2}\right],
\ee
where $l_{\rm st}$ is the string length scale, $\hat T(k):=T(k)/T_{\rm H}$ and the temperature $T(k)$ is evaluated at the time when the mode with comoving wave-number $k$ exits the horizon. As said in the introduction, the form of $T(k)$ is unknown except for its behaviour during the Hagedorn phase ($T\approx {\rm const}\lesssim T_{\rm H}$) and in the following radiation-domination era ($T\sim 1/a$). 

Although we have the full expression for the tensor power spectrum in terms of $\hat{T}(k)$, we do not have the correspondence between $\hat{T}(k)$ and $k$. Thus, we do not have a way to plot the GW power spectrum as a function of $k$ using \eqref{sgc2}. Here, relying on the approximation that the SGWB is generated when $T$ is almost constant during the Hagedorn phase, we plot the GW spectrum using the power series expansion \Eq{Ptk} given in terms of $n_{\rm t}$ and $\a_{\rm t}$. However, as we will discuss below, this expression can break down at high frequencies. To show that, we will compare \Eq{Ptk} with the largest amplitude of GWs obtained from the full expression. 

%However, this is sufficient to establish the primordial spectrum, which is generated during the Hagedorn phase where $T$ is almost constant and one can use the parametrization \Eq{Ptk}. 
Let us first briefly describe the CMB observables. Since the spectrum is almost scale invariant, the tensor tilt is small but, contrary to the standard inflationary paradigm, positive. In deriving $n_{\rm t}$, usually both the logarithmic term in \Eq{sgc2} and the running of the spectral index are ignored \cite{BrNP}, but the SGBW is sensitive to small variations of the spectral index and its running. Here we will retain the full expressions of $n_{\rm t}$ and the newly derived $\a_{\rm t}$ before approximating them.
%commenting on the validity of this approximation in section \ref{ressgc}. 

Since the scalar power spectrum is $\cP_{\rm s}\propto \hat T/(1-\hat T)$, the exact expression of the scalar spectral index is
\be\label{sgcns}
n_{\rm s}-1 = \frac{\rmd\ln\cP_{\rm s}}{\rmd\ln k}= \frac{1}{\hat T(1-\hat T)}\frac{\rmd \hat T}{\rmd\ln k}\,,
\ee
where the factor $1/\hat T$ is usually approximated to 1 \cite{BrNP}. Since $\hat T\lesssim 1$ and $\hat T$ decreases with $k$ ($\rmd \hat T/\rmd\ln k<0$), an increase of power is observed for small $k$ ($n_{\rm s}-1<0$). A small red tilt in the scalar sector is produced because scalar modes are generated by the energy density, which increases with $T$. The running of the scalar index is negative, as one can appreciate from the approximate consistency relation $\a_{\rm s}\simeq-(1-n_{\rm s})$ \cite{Brandenberger:2017xjy}.

Using \Eq{sgcns}, we find
\ba
n_{\rm t}&=& \frac{\rmd\ln \cP_{\rm t}}{\rmd\ln k}=(1-n_{\rm s})(2\hat T-1)-\frac{2[2-(1-n_{\rm s})\hat T]}{\ln[(1-\hat T)(l_{\rm st}k)^{-2}]}\nonumber\\
&\simeq& (1-n_{\rm s})-\frac{4}{\ln[(1-\hat T)(l_{\rm st}k)^{-2}]}>0\,.
\label{ntstga}
\ea
From the first line of \Eq{ntstga}, we obtain
\ba
\a_{\rm t}&=& -\a_{\rm s}(2\hat T-1)-2(1-n_{\rm s})^2\hat T(1-\hat T)\nonumber\\
&&-\frac{2\hat T[\a_{\rm s}+(1-n_{\rm s})^2(1-\hat T)]}{\ln[(1-\hat T)(l_{\rm st}k)^{-2}]}-\frac{2[2-(1-n_{\rm s})\hat T]^2}{\ln^2[(1-\hat T)(l_{\rm st}k)^{-2}]}\nonumber\\
%-\a_{\rm s}\left\{2\hat T-1+\frac{2\hat T}{\ln[(1-\hat T)(l_{\rm st}k)^{-2}]}\right\}\nonumber\\
%&&-2(1-n_{\rm s})^2\hat T(1-\hat T)\left\{1+\frac{1}{\ln[(1-\hat T)(l_{\rm st}k)^{-2}]}\right\}-\frac{2[2-(1-n_{\rm s})\hat T]^2}{\ln^2[(1-\hat T)(l_{\rm st}k)^{-2}]}\nonumber\\
&\simeq&-\a_{\rm s}\,,
\label{atstga}
\ea
where in the last line we dropped higher-order terms in the $\hat T\sim 1$ expansion. Therefore, the running of the tensor index has opposite sign with respect to the scalar running.

A blue-tilted tensor spectrum is one of the characteristic predictions of string-gas cosmology that could be tested if a primordial gravitational signal was discovered. However, in order to get a detectable signal, the small blue tilt should be accompanied by as high as possible an $r$. The tensor-to-scalar ratio depends on the string scale $l_{\rm st}$:
\be
r=\frac{9}{4}\left(1-\hat T\right)^2\ln^2\left[\frac{1-\hat T}{(l_{\rm st}k)^2}\right]\,,
\label{rstga}
\ee
where the additional pre-factor $9/4$ compared to \cite{BrNPV,BNPV,BrNP} comes from the fact that we define the tensor-to-scalar ratio in terms of the two-point function of the comoving curvature perturbation $\cP_{\rm s}\propto \langle\cR^2\rangle$, while \cite{BrNPV,BNPV,BrNP} use the that of the scalar perturbation $\langle\Phi^2\rangle$, which is related to the curvature perturbation $\zeta\simeq \cR$ by $\Phi\simeq -(2/3)\zeta$ for a radiation dominated era.

When $l_{\rm st}\sim 10^{-3}\lpl$ and $\hat T\approx 0.99$, one obtains an observable $r\lesssim O(0.1)$ \cite{BrNP} while respecting non-Gaussianity bounds \cite{CWXB}.

\begin{figure}
\centering
\includegraphics[width=10cm]{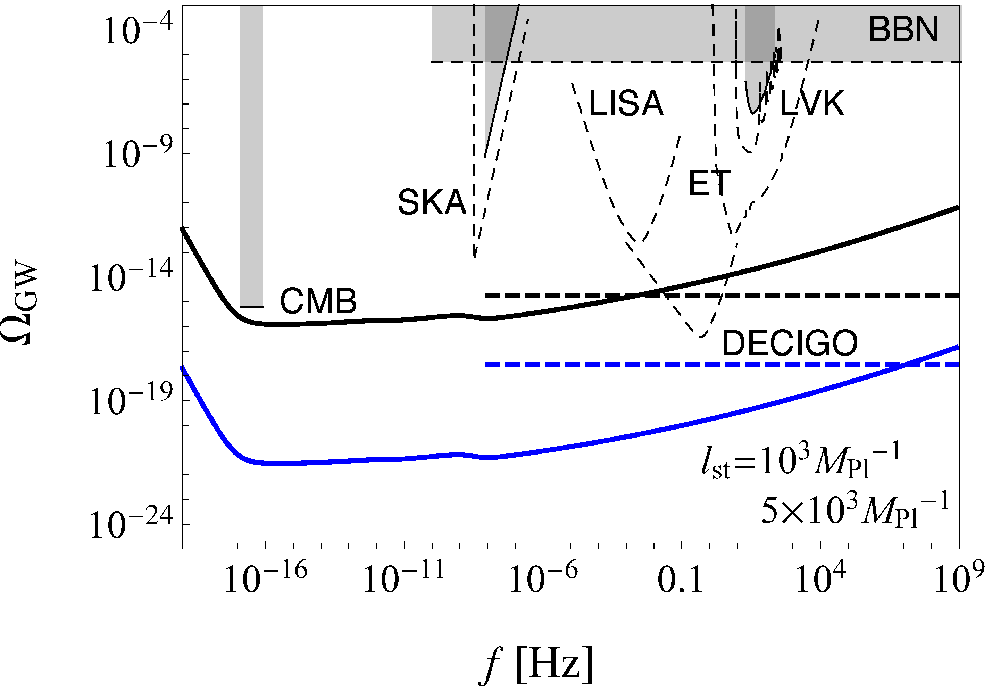}
\caption{\label{fig7} 
SGWB of string-gas cosmology compared with the sensitivity curves of LIGO-Virgo-KAGRA (LVK), SKA, LISA, ET and DECIGO. 
%Denoting as $n_{\rm s}^{\rm obs}\pm\de n_{\rm s}$ and $\a_{\rm s}^{\rm obs}\pm\de \a_{\rm s}$ the \textsc{Planck} values \Eq{obspl}, we plot the worse case minimizing the tensor blue tilt and maximizing the negative tensor running at the $2\s$-level ($n_{\rm t}=1-(n_{\rm s}^{\rm obs}+2\de n_{\rm s})\approx 0.026$, $\a_{\rm t}=-(\a_{\rm s}^{\rm obs}+2\de \a_{\rm s})\approx -0.007$, red solid curve), the intermediate case taking the central values of the parameters ($n_{\rm t}=1-n_{\rm s}^{\rm obs}\approx 0.034$, $\a_{\rm t}=-\a_{\rm s}^{\rm obs}\approx 0.007$, black solid curve) and the best case maximizing the tensor blue tilt and the positive tensor running at the $2\s$-level ($n_{\rm t}=1-(n_{\rm s}^{\rm obs}-2\de n_{\rm s})\approx 0.042$, $\a_{\rm t}=-(\a_{\rm s}^{\rm obs}-2\de \a_{\rm s})\approx 0.021$, blue solid curve). The dashed curves correspond to the above cases at the $1\s$-level.
The black and blue solid curves corresponds to $l_{\rm st}=10^3 \Mpl^{-1}$ and $5\times 10^3 \Mpl^{-1}$, respectively. We used the central values of the \textsc{Planck} constraints \Eq{obspl}: $n_{\rm s}=0.9658$ and $\a_{\rm s}=-0.0066$. Note that the approximation \Eq{Ptk} fails for high-frequency GWs, since the corresponding curves go beyond the upper bound on the GW amplitude (dashed lines) obtained from the full expression \eqref{sgPTmax}.}
\end{figure}

Using the power-series parametrization \Eq{Ptk} and the CMB predictions \eqref{ntstga}-\eqref{rstga}, we plot the spectrum of the SGWB of string-gas cosmology in Fig.\ \ref{fig7}. As seen in \eqref{ntstga} and \eqref{atstga}, $n_{\rm t}$ and $\a_{\rm t}$ can be expressed in terms of scalar observables $n_{\rm s}$ and $\a_{\rm s}$. We use the central values of the \textsc{Planck} constraints \Eq{obspl}, $n_{\rm s}=0.9658$ and $\a_{\rm s}=-0.0066$. We find that the SGWB can reach the sensitivity of DECIGO if the string scale is $l_{\rm st} = 10^3 \Mpl^{-1}$ or smaller. However, these curves fail to capture the full spectrum of the model, also because the \textsc{Planck} central values do not respect the consistency relation $\a_{\rm s}\simeq-(1-n_{\rm s})$ \cite{Brandenberger:2017xjy}.

In the figure, we also show the upper bound on the GW amplitude obtained from the full expression, which can be obtained as follows. From \eqref{sgc2}, we find that the function takes the maximum value at $\hat T(k)\approx 0.5$ when $l_{\rm st}^2k^2 \ll 1$, which is the case for the GW frequencies of our interest, since $l_{\rm st}\sim (3.7\times 10^{39}\,{\rm Hz})^{-1}$ for $l_{\rm st}=10^3 \Mpl^{-1}$. The maximum value is 
\be\label{sgPTmax}
\cP_{\rm t,max}\simeq\frac{0.0625}{(\Mpl l_{\rm st})^4}\,\ln^2\left[\frac{0.5}{l_{\rm st}^2k^2}\right]\,.
\ee
For example, for $l_{\rm st}=10^3 \Mpl^{-1}$, we find that the function takes the maximum value at $\hat T(k)=0.496$ and $\cP_{\rm t,max}=1.89\times 10^{-12}$. The upper bounds are plotted as thick dashed lines in the figure and they clearly show that the power-series expansion extrapolated from the CMB scale overestimates the GW amplitude at high frequencies. 

%In the code, we used the exact expressions for $n_{\rm t}$ and $\a_{\rm t}$ but we checked \emph{a posteriori} that the lowest-order approximated expressions $n_{\rm t}\simeq 1-n_{\rm s}$ and $\a_{\rm t}\simeq -\a_{\rm s}$ are sufficient. The reason to use the full expressions is that the logarithmic term in the tensor spectrum gives a sizable contribution (about $50\,\%$ of its value) to $n_{\rm t}$ but not to $\a_{\rm t}$. At the pivot scale \Eq{pivot} and from equation \Eq{lpmpc}, for a grand-unification string scale $l_{\rm st}=(10^{16}\,{\rm GeV})^{-1}\approx 5\times 10^{-55}\,{\rm Mpc}$ and $\hat T\approx 0.99$, the second term in \Eq{ntstga} is $\approx 0.016$, to be compared with the first term $\approx 0.035$. Therefore, the tensor spectrum \Eq{ntstga} is blue-tilted but less than if we had taken the crudest approximation $n_{\rm t}\simeq 1-n_{\rm s}$.

%However, while the logarithmic term omitted in the relation $n_{\rm t}\simeq 1-n_{\rm s}$ does change the value of the tensor index considerably, the difference is small compared to the amplitude scale of the figure and the phenomenology is basically the same. This happens because the main agent tilting the spectrum to the blue is the running of the tensor index, not the index itself. On the other hand, the terms ignored in $\a_{\rm t}\simeq -\a_{\rm s}$ give a small correction to the value of the running, so that overall the lowest-order approximation turns out to be adequate to study the predictions of the model.

%%%%%%%%%%%%%%%%%%%%%%%%%%%%%%%%%%%%%%%%%%%%%%%%%%%%%%%%%%%%%%%%%%%%%%%%%%%%%%%%%%%%%%%%%%%%%%%%%%%%

\subsection{New ekpyrotic scenario}

In the ekpyrotic universe, two flat 3-branes constitute the boundary of a five-dimensional spacetime and interact with an attractive potential $V(\vp)$ along a compact fifth dimension parametrized by the radion $\vp$. These branes can be the orbifold planes of heterotic M-theory (M-theory compactified on $\cC_3\times S^1/\mathbbm{Z}_2$, where $\cC_3$ is a Calabi--Yau space) \cite{HoWi1,HoWi2} or the $(3+1)$-dimensional manifolds of a Randall--Sundrum setting \cite{RaSu2}. As the branes get closer, the gravitational energy in the bulk is converted into brane kinetic energy. Since the branes are boundary ones, instead of collapsing via tachyon condensation they collide and oscillate back and forth their center of mass along the extra direction. During the collision at coincident branes ($\vp=-\infty$), part of the brane kinetic energy is converted into matter and radiation. An observer on one of the branes experiences the brane collision as a big bang (vanishing scale factor $a_{\rm E}$ in the Einstein frame) after a period of contraction called ekpyrosis. Even if the fifth dimension experiences a big crunch and it collapses to a point, and even if $a_{\rm E}(t)=0$ for the brane observer, the brane metric in the Jordan frame, the local temperature and the energy density on the brane remain finite at the event. 

During the slow contraction phase, a pattern of inhomogeneities is developed. In fact, due to quantum fluctuations the branes are not parallel at all points and they collide at slightly different times in different places. These patches begin their evolution and cooling down from the bounce out of sync, which causes the anisotropies observed in the sky. The spectral properties of this pattern depend on the specific model of ekpyrosis (see \cite[chapter 11]{CQC} for a review). Here we concentrate on the most recent one, of which we already gave a bird's eye view in the introduction; we refer to \cite{Brandenberger:2020tcr,Brandenberger:2020eyf} for more details. The novelty with respect to old models is the inclusion of an S-brane at high energies, which is expected to be there on theoretical grounds. If the S-brane has vanishing shear, one obtains a viable cosmology with the consistency relation \cite{Brandenberger:2020eyf}
\be
n_{\rm t}=1-n_{\rm s}>0\,,\label{ntekpy}
\ee
where the scalar spectrum is red-tilted. Therefore,
\be
\a_{\rm t}=-\a_{\rm s}\,.
\ee
The tensor-to-scalar ratio at the pivot scale is\footnote{Note that, for the same reason described in string-gas cosmology, we have the additional factor $9/4$ compared to \cite{Brandenberger:2020eyf}.}
\be\label{rekpy}
r(k_0)\simeq \frac{9}{4}\frac{9\cdot 2^{n_{\rm s}}}{\Gamma^2\!\left(1-\frac{n_{\rm s}}{2}\right)}(k_0\tau_{\rm B})^{2(1-n_{\rm s})}\,(1-n_{\rm s})^2\,,
\ee
where $\Gamma$ is Euler's function and $\tau_{\rm B}$ is the conformal time at which the string density is reached and the cosmological bounce takes place. Taking the grand-unification scale at $10^{16}$\,GeV, we obtain $\tau_{\rm B}=(aH)^{-1}\approx 5.8\times 10^{-24}\,{\rm Mpc}$. With the observed scalar index \Eq{obsns}, one has $r(k_0)\approx (2\mhyp 5) \times 10^{-4}$.

\begin{figure}
\centering
\includegraphics[width=10cm]{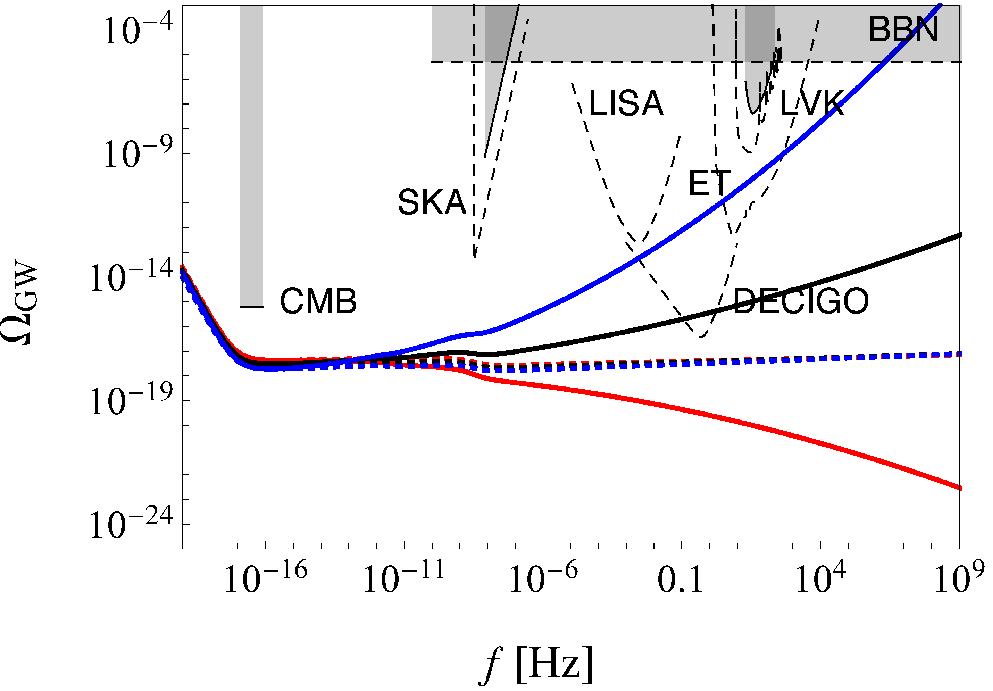}
\caption{\label{fig8} SGWB of the new ekpyrotic scenario %with $\cN=50$
 compared with the sensitivity curves of LIGO-Virgo-KAGRA (LVK), SKA, LISA, ET and DECIGO. Denoting as $n_{\rm s}^{\rm obs}\pm\de n_{\rm s}$ and $\a_{\rm s}^{\rm obs}\pm\de \a_{\rm s}$ the \textsc{Planck} values \Eq{obspl}, we plot the worst case minimizing the tensor blue tilt and maximizing the negative tensor running at the $2\s$-level ($n_{\rm t}=1-(n_{\rm s}^{\rm obs}+2\de n_{\rm s})\approx 0.026$, $\a_{\rm t}=-(\a_{\rm s}^{\rm obs}+2\de \a_{\rm s})\approx -0.007$, red solid curve), the intermediate case taking the central values of the parameters ($n_{\rm t}=1-n_{\rm s}^{\rm obs}\approx 0.034$, $\a_{\rm t}=-\a_{\rm s}^{\rm obs}\approx 0.007$, black solid curve) and the best case maximizing the tensor blue tilt and the positive tensor running at the $2\s$-level ($n_{\rm t}=1-(n_{\rm s}^{\rm obs}-2\de n_{\rm s})\approx 0.042$, $\a_{\rm t}=-(\a_{\rm s}^{\rm obs}-2\de \a_{\rm s})\approx 0.021$, blue solid curve). We take $\tau_{\rm B} = 5.8\times 10^{-24}\,{\rm Mpc}$, which corresponds to the conformal time at the grand-unification scale $10^{16}$\,GeV. The dotted curves correspond to the above cases with no running.}
\end{figure}

Figure \ref{fig8} shows the SGWB of this model, obtained using the power-series approximation \Eq{Ptk}. The parameters $r$, $n_{\rm t}$ and $\a_{\rm t}$ can be expressed in terms of the scalar observables $n_{\rm s}$ and $\a_{\rm s}$ and we take $\tau_{\rm B}=5.8\times 10^{-24}\,{\rm Mpc}$. In the figure, we show the case where we use the central values of the \textsc{Planck} constraints \Eq{obspl} as well as the cases of minimizing/maximizing the tensor spectrum (largest/smallest values of $n_{\rm s}$ and $\a_{\rm s}$ at the $2\s$ level). We also show the spectra calculated assuming no tensor running (dotted lines), which differ a lot with respect to the case where we take the running into account. This indicates the importance of the inclusion of the running for predicting the amplitude of high-frequency GWs. We find that, in the most optimistic case with bluest tilt and strongest positive running (at $2\s$ in the experimental error), the signal reaches the sensitivity curves of ET and DECIGO.

Note that the evolution during the ekpyrotic phase has not been investigated in detail and, thus, the behaviour of the GW spectrum at high frequency could deviate from the prediction obtained from the power-series parametrization, as was the case for non-local Starobinsky inflation and string gas cosmology. Therefore, the figure only presents a rough estimate on the SGWB amplitude at interferometer scales, inferred from the current constraints of CMB observables. 

%%%%%%%%%%%%%%%%%%%%%%%%%%%%%%%%%%%%%%%%%%%%%%%%%%%%%%%%%%%%%%%%%%%%%%%%%%%%%%%%%%%%%%%%%%%%%%%%%%%%

\subsection{Brandenberger--Ho non-commutative inflation}

In this model, time and space coordinates do not commute, $[\tilde\tau,x]=\rmi/M^2$, where $\tilde\tau:=\int\rmd t\,a$ and $x$ is a comoving spatial coordinate. The fundamental mass scale $M$ defines the correction parameter
\be
\delta := \left(\frac{M}{H}\right)^2
\ee
and roughly divides the space of comoving wave-numbers into two regions: a nearly commutative one where the Hubble radius at inflation was larger than the fundamental length ($H\ll M$, $\de\gg 1$) and perturbations were generated inside the horizon (thus, $k_{\rm p}=k/a\ll M$, where $k_{\rm p}=1/\la$ is the proper frequency), and another describing the IR, large-scale perturbations created outside the sub-Planckian horizon ($H\gg M$, $\de\ll 1$) \cite{BH}. The IR limit is more speculative but it has the largest non-commutative effects and it admits inflationary potentials giving rise to observables compatible with data \cite{Calcagni:2013lya}. In this case, and ignoring the UV limit from now on, the spectrum of tensor perturbations (the scalar spectrum has the same shape) is
\be\label{Anoncom}
\cP_{\rm t} = \cP^{(0)}_{\rm t}\,\Sigma^2 (\delta)\,,
\ee
where $\cP_{\rm t}^{(0)}=\cP_{\rm t}(\Sigma\!\!=\!\!1)$ is the amplitude in the commutative limit (Einstein gravity) and $\Sigma(\delta)$ is a function encoding the non-commutative effects. The factor $\Sigma$ multiplies both the tensor and scalar amplitudes, so that their ratio $r$ is the usual one when expressed in terms of $\e$.

To lowest order in the slow-roll parameters, one has $\rmd \ln \Sigma^2/\rmd\ln k\simeq\sigma \e$, where $\s$ is a function of $\delta$ such that $\dot\s=O(\e)$ \cite{Cal4}. In the commutative case, $\sigma=0$. The IR model admits two values, $\s=2$ and $\s=6$, depending on how non-commutativity enters the second-order action for the cosmological perturbations \cite{Cal4}. For instance, in the $\s=6$ case, the FLRW 2-sphere is factored out of the total measure, while in the other case it is not. The tensor spectral index is given by
\be\label{ntnc}
n_{\rm t} \simeq -(2-\s)\e\,.
\ee
The scalar index $n_{\rm s}-1$ gets the same correction of $+\s\e$. Since we are interested in blue-tilted tensor spectra, we will focus on $\s=6$, although we will recall the general formul\ae\ of the model. The tensor-to-scalar ratio reads
\be\label{ratio}
r\simeq 16\e\simeq-\frac{16}{2-\s}n_{\rm t}\,,
\ee
while the runnings of the tensor spectral index is
\be
\a_{\rm t} \simeq -2(2-\s)\e(\e-\eta)\,,\label{alphat}
\ee
for constant $\s$. In particular, for $\s=6$, we find
\be
n_{\rm t}\simeq \frac{r}{4}\simeq 4\e,\qquad \a_{\rm t} \simeq 8\e(\e-\eta)\,.\label{nt0}
\ee

When $\s=6$, the tensor spectrum becomes automatically blue-tilted but the scalar spectrum can stay red-tilted for some choices of inflaton potential \cite{Calcagni:2013lya}. A viable inflationary model embedded in this non-commutative setting with red-tilted scalar spectrum is natural inflation \cite{FFO,ABFFO}, characterized by the potential 
\be
V(\phi)=V_0 \left(1+\cos\frac{\phi}{\phi_*}\right),
\label{napotential}
\ee
where $V_0>0$ and $\phi_*$ is the energy scale at which the global symmetry 
associated with this model is broken. The field is related to the number of e-foldings by
\be
x:=\cos\left(\frac{\phi}{\phi_*}\right)=1-\frac{12A}{1+6A}\rme^{-\frac{\cN}{3A}}\,,
\label{xna}
\ee
where $A=(\phi_*/\Mpl)^2/3$. Since 
\be
\e\simeq \e_V=\frac{\Mpl^2}{2}\left(\frac{V_{,\phi}}{V}\right)^2=\frac{1}{6A}\frac{1-x}{1+x}=\frac{1}{(1+6A)\,\rme^{\frac{\cN}{3A}}-6A}\,,
\ee
one has \cite{Calcagni:2013lya}
\be
r \simeq \frac{8}{3A} \frac{1-x}{1+x}\,,\qquad n_{\rm t}\simeq \frac{2}{3A} \frac{1-x}{1+x}\,.\label{nanon}
\ee
The large-$A$ limit is excluded by observations, while small or intermediate values of $A$ lie within the likelihood region of \textsc{Planck}+WP+BAO+high-$\ell$ data. For $\s=6$ and choosing the pivot wave-number \Eq{pivot}, one has \cite{Calcagni:2013lya}
\be\label{boundA}
5.8<A<11 \quad (95\,\%\,{\rm CL}) \quad {\rm for} \quad \cN=50\mhyp 60\,.
\ee
From these expressions, one can see that the typical predictions of this model are $n_{\rm t} =0.004\mhyp 0.015$ and $r=0.02\mhyp 0.06$. The tensor running is positive, since $\e$ decreases with the number of e-foldings:
\be
\a_{\rm t} = -4 \frac{\rmd\e}{\rmd\cN}\simeq\frac{4}{9A^2}\frac{1-x}{(1+x)^2}>0\,.
\ee

\begin{figure}
\centering
\includegraphics[width=10cm]{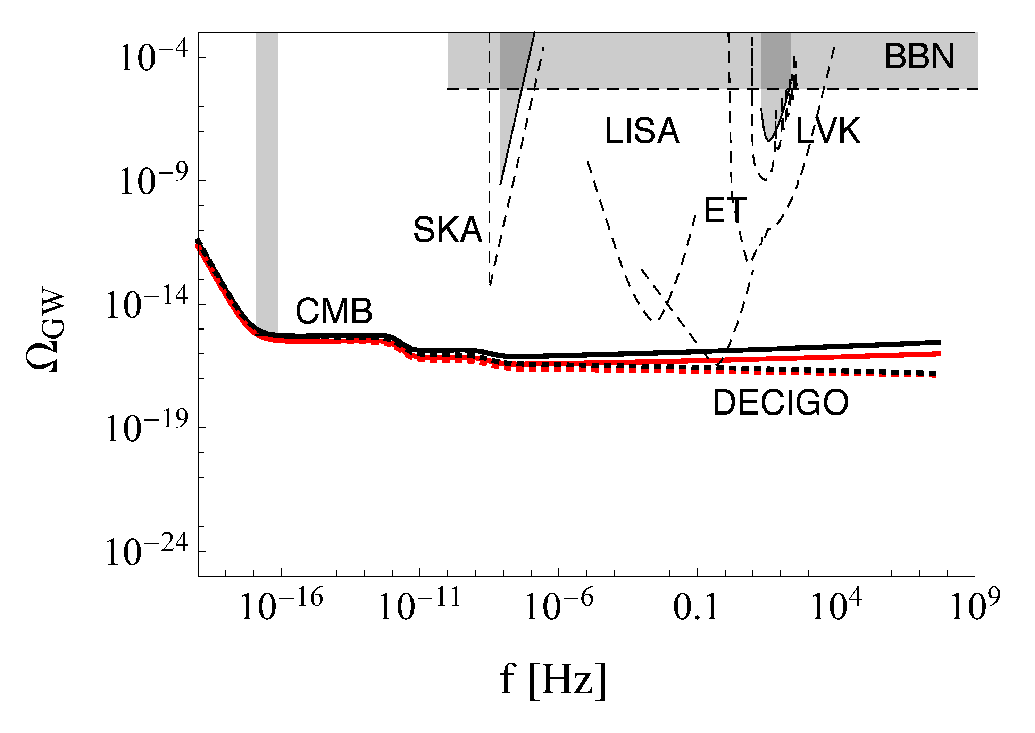}
\caption{\label{fig9} SGWB of the Brandenberger--Ho non-commutative model with $\cN=50$ compared with the sensitivity curves of LIGO-Virgo-KAGRA (LVK), SKA, LISA, ET and DECIGO. The red and black solid curves represent the cases with, respectively, $A=7$ and $A=10$, where $A$ is given in \Eq{xna} and the values lie within the allowed interval \Eq{boundA}. Dotted curves correspond to the commutative cases ($\s=0$) with the above values of $A$. Note that we widened the single-frequency CMB constraint to a band in order to make it more visible and that the spectra in the plot satisfy the CMB bound at the pivot scale.}
\end{figure}

The SGWB of this model is depicted in Fig.\ \ref{fig9} when taking into account up to the second order ($\alpha_t$) in the power-series expansion \Eq{Ptk}. As expected by the fact that non-commutativ\-i\-ty changes the sign and value of the coefficients in the slow-roll expressions of the observables but not their order of magnitude, the spectrum is enhanced up to the DECIGO sensitivity curve, but barely so, and it does not reach any other interferometer experiment.

%%%%%%%%%%%%%%%%%%%%%%%%%%%%%%%%%%%%%%%%%%%%%%%%%%%%%%%%%%%%%%%%%%%%%%%%%%%%%%%%%%%%%%%%%%%%%%%%%%%%

\subsection{Multi-fractional inflation}

Multi-fractional spacetimes are spacetimes where the clocks and rulers used by the observer register different scaling laws (for instance, linear size versus volume) in copies of the same object with  different sizes (for instance, a human-size ball compared with a microscopic one) \cite{revmu}. This ever-changing geometry is typical of spacetimes arising in quantum gravity \cite{tH93,Car09,fra1,Car17}. Multi-fractional theories implement it via a modification of the integration measure in the field action and of the kinetic operators acting on the fields, including gravity. While the integral structure is changed according to basic principles of multi-scaling and fractal geometry into a unique parametrization, there are different ways to deform the differential structure. The main ones make use of weighted derivatives, $q$-derivatives or fractional derivatives \cite{revmu}.

Inflation in multi-fractional spacetimes has been studied in the so-called theory with $q$-derivatives \cite{frc14}. Since we will mainly work in momentum space, we do not have to recall what $q$-derivatives in position space are. Suffice it to say that the gravitational sector of the theory looks like Einstein gravity but with coordinates $x^\mu$ replaced by fixed profiles $q^\mu(x^\mu)$ which break ordinary diffeomorphism invariance. These composite coordinates depend on the fundamental lengths of the geometry and also display a discrete scaling symmetry encoded in logarithmic oscillations. Ignoring these oscillations, assuming that geometry has only one fundamental scale and considering an isotropic configuration where all spatial directions scale in the same way, the composite spatial comoving momentum coordinates associated with the $q_i(x^i)$ are
\be\label{pk}
p_i(k^i) \simeq k^i\left[1+\frac{1}{|\a|}\left(\frac{k^i}{k_*}\right)^{1-\a}\right]^{-1},
\ee
where $k_*$ is a fundamental comoving scale and the parameter $\a$ can take one of the following ranges of values:
\ba
\textrm{case 1}: &&\qquad\!\a<0\qquad\textrm{(UV modification of gravity)}\,,\\
\textrm{case 2}: &&0<\a<1\qquad\textrm{(UV modification of gravity)}\,,\\
\textrm{case 3}: &&\qquad\!\a>1\qquad\textrm{(IR modification of gravity)}\,.
\ea
An expression similar to \Eq{pk} holds also for the time direction, with fractional parameter $\a_0$ not necessarily equal to $\a$. The Hausdorff dimension of spacetime is the sum of the $\a$ parameters, $\dh=\sum_{\mu=0}^{D-1}\a_\mu=\a_0+(D-1)\a$. Therefore, assuming $\a_0=O(1)$, in $D=4$ topological dimensions spacetime has a smaller dimensionality $\dh<4$ in the UV in cases 1 and 2 and a larger dimensionality $\dh>4$ in the IR in case 3. In case 1, we assume that the Hausdorff dimension is well defined, i.e.,
\be\label{valal}
\dh>0\qquad\Longrightarrow\qquad \a_0>|3\a|\,.
\ee
Cases 1, 2 and 3 are not mutually exclusive. If one takes a three-term generalized polynomial for $p_i$ with three different scales $k_{*123}$, one obtains a geometry where the Hausdorff dimension changes in the UV to two different values $\dh<4$, $\dh=4$ at intermediate scales (general-relativity regime) and an IR regime where $\dh>4$. Removing regimes 1 and 2 from the picture, one is left with pure IR multiple modifications of gravity, while removing either regime 1 or regime 2 one ends up with a model with only one exotic UV regime. From now on, we will consider only geometries with one exotic regime (case 1 or 2 or 3). We will call this exotic regime fractional.

\begin{figure}
\centering
\includegraphics[width=10cm]{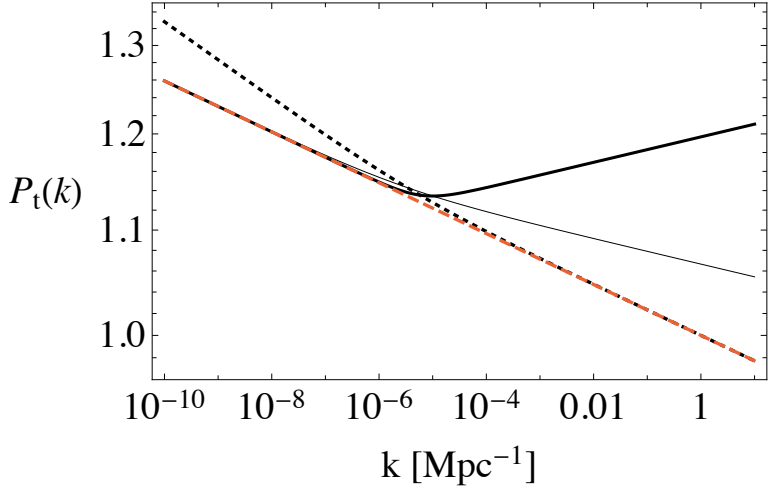}
\caption{\label{fig10} Primordial tensor spectrum \Eq{Pt} for $\a<0$ ($\a=-1/2$ thick curve), $0<\a<1$ ($\a=1/2$, thin curve) or $\a>1$ ($\a=3/2$, dotted curve), with $k_*=10^{-5}$\,Mpc$^{-1}$ and $n_{\rm t}=-0.01$. The case of Einstein gravity is the dashed line.}
\end{figure}

Inflation is driven by a scalar field slowly rolling down its potential. In terms of the $q$ coordinates, the inflationary dynamics is the same as in general relativity, and so are the power spectra in terms of the $p$ momenta. In particular, for an isotropic configuration one can write an approximate expression for $p:=\sqrt{p_1^2(k^1)+p_2^2(k^2)+p_3^2(k^3)}$ in terms of the absolute value $|\bm{k}|=\sqrt{k_1^2+k_2^2+k_3^2}$. This is nothing but equation \Eq{pk} where $k=|\bm{k}|$. Since the standard slow-roll approximation is valid in the frame described by $q$ and $p$ coordinates, instead of the parametrization \Eq{Ptk} one should use
\be\label{Pt}
\cP_{\rm t}(k)=\cP_{\rm t}(k_0)\,\exp\left\{n_{\rm t}(k_0)\,\ln\frac{p(k)}{p(k_0)}+\frac{\a_{\rm t}(k_0)}{2}\left[\ln\frac{p(k)}{p(k_0)}\right]^2\right\},\qquad \cP_{\rm t}(k_0)=r(k_0)\,\cP_{\rm s}(k_0)\,,
\ee
where $n_{\rm t}<0$ and $\a_{\rm t}<0$ as in standard inflationary models (equation \Eq{eingrav}) and
\be
p(k) \simeq k\left[1+\frac{1}{|\a|}\left(\frac{k}{k_*}\right)^{1-\a}\right]^{-1}.\label{Pt2}
\ee
The consistency relation between the tensor-to-scalar ratio and the tensor index is the same as in general relativity,
\be\label{r8n}
r=-8n_{\rm t}\,.
\ee
We can always identify two regimes whose $k$-range depends on the sign and value of $\a$:
\ba
\textrm{General-relativity regime:} &\qquad p(k)\simeq k\,,\qquad & P_{\rm t}(k)\sim k^{n_{\rm t}+\frac12\a_{\rm t}\ln(k/k_0)}\,,\label{grreg}\\
\textrm{Fractional regime:} &\qquad p(k)\simeq k^\a\,,\qquad & P_{\rm t}(k)\sim k^{\a n_{\rm t}+\frac12\a_{\rm t}\a^2\ln(k/k_0)}\,.\label{frareg}
\ea
We can distinguish three cases (Fig.\ \ref{fig10}).
\begin{enumerate}
\item $\a<0$. This is the most interesting case for a SGWB. For $k\ll k_*$ (large scales, small frequencies), one hits the general-relativity regime, while the fractional regime is reached for $k\gg k_*$ (small scales, large frequencies). In the fractional regime, the effective spectral index $\a n_{\rm t}$ is positive definite and the spectrum increases at large $k$. From \Eq{frareg}, the effective tensor running at large frequencies stays negative, $\a^2\a_{\rm t}<0$, and its absolute value is larger than in standard cosmology if $\a<-1$. Intuitively, the enhancement of the power spectrum at small scales compared to Einstein gravity occurs because at these scales $\dh<4$ but the primordial density is the same at all scales. Therefore, the number of small-scale modes slightly increases while the number of large-scale modes is the same as standard.  Hence the spectrum is \emph{blue-tilted} but with negative running.
\item $0<\a<1$. The general-relativity regime \Eq{grreg} corresponds to $k\ll k_*$ (large scales, small frequencies), while the fractional regime \Eq{frareg} corresponds to $k\gg k_*$ (small scales, large frequencies), just like in case 1. The effective spectral index $\a n_{\rm t}$ is negative and smaller than usual due to the suppression factor $\a$. Because of \Eq{frareg}, also the running is negative and, in particular, at large frequencies $\a^2 \a_{\rm t}<0$. This case is the most studied in the literature \cite{frc14,frc11} and corresponds to a theory where geometry is modified in the UV. The geometric interpretation is that in the UV $\dh<4$ and there is an increase in the number of modes given the same density. Overall, the tensor spectrum is \emph{less red-tilted} than in Einstein gravity.
\item $\a>1$. The above limits are switched but the result is similar: the effective spectral index $\a n_{\rm t}$ is negative, although this time it is larger than usual due to the enhancement factor $\a$. The running is negative and, in particular, at large frequencies $\a^2|\a_{\rm t}|$ is larger than in standard cosmology. Again, the geometric interpretation is intuitive: $\dh>4$ at large scales or small frequencies and tensor modes have ``more spacetime'' than in Einstein gravity. Their density is constant, which means that their number decreases at large scales (suppressed spectrum). Since the spectrum is standard at small scales, overall the tensor spectrum is \emph{more red-tilted}.
\end{enumerate}

Note that the parameters and $k$-dependence of the scalar spectrum are the same (with $n_{\rm t}$ replaced by $n_{\rm s}-1$), so that if the tensor spectrum is blue-tilted, so is the scalar one. This does not constitute a problem if the transition scale $k_*$ is large enough, in which case both spectra are red-tilted at the largest inflationary scales $k\lesssim k_0$ and the scalar sector respects observational constraints.

Another remark concerns the fact that, as in the rest of the literature, we defined the model in terms of a comoving constant fundamental scale $k_*$ rather than a constant fundamental proper scale $k_{{\rm p}*}=k_*/a$, as done in the non-commutative model. However, doing so would lead to a model without observable effects in the SGWB. Consider a power-law expansion $a=t^{1/\e}$, so that conformal time is $\tau=\int\rmd t/a\propto t^{1-1/\e}$ and $t\propto \tau^{\e/(\e-1)}$. At horizon crossing ($\tau=1/k$), one has $a^{-1}\propto k^{-1/(1-\e)}\simeq k^{-1-\e}$ during inflation. Instead of the ratio $(k/k_*)^{1-\a}$ in the spectrum \Eq{Pt}, we would have a physical scale $k_{{\rm p}*}$ and the ratio $[k/(a k_{{\rm p}*})]^{1-\a}=(H/k_{{\rm p}*})^{1-\a}$, so that the discrimination between UV and IR regimes would be set by the value of the Hubble horizon during inflation (respectively, $H\ll k_{{\rm p}*}$ and $H\gg k_{{\rm p}*}$). Therefore, inflation would take place only in one of these two regimes and all the scales exiting the horizon would be either in the UV or in the IR. However, the frequencies of all present and future GW interferometers (including DECIGO) are not very high and correspond to scales that exited the horizon during inflation. In particular, a blue tilt at the DECIGO frequency scale would imply a blue tilt also during inflation, including in the scalar sector. Contrary to the non-commutative model, the scalar spectrum is guaranteed to be blue-tilted if the tensor spectrum is such, since $P_{\rm s}(k)\sim k^{\a(n_{\rm s}-1)}$ and $n_{\rm s}-1<0$. This model would then be unviable.
%Another remark concerns the fact that we defined the model in terms of a comoving constant fundamental scale $k_*$ rather than a physical constant fundamental scale $E_*$, as done in the non-commutative model. However, doing so would not change the overall picture and, in fact, would complicate it. Consider a power-law expansion $a=t^{1/\e}$, so that conformal time is $\tau:=\int^t\rmd t/a\propto t^{1-1/\e}$ and $t\propto \tau^{\e/(\e-1)}$. At horizon crossing ($\tau=1/k$), one has $a^{-1}\propto k^{-1/(1-\e)}\simeq k^{-1-\e}$ during inflation. Instead of the ratio $(k/k_*)^{1-\a}$ in the spectrum \Eq{Pt} we would have a physical scale $E_*$ and the ratio $[k/(a E_*)]^{1-\a} \sim (k/E_*)^{-\e(1-\a)}=:(k/E_*)^{1-\tilde\a}$, where $\tilde\a = 1+\e(1-\a)$. Everything said above would hold for $\tilde\a$ instead of $\a$. In particular, for $\e>0$ (ordinary inflationary scenarios), case 1 ($0<\tilde\a<1$) corresponds to $\a>1$, case 2 ($\tilde\a>1$) to $\a<1$, including negative values, and case 3 ($\tilde\a<0$) to $\a\gg 1$. Although these new parameter ranges would change the geometric interpretation of the three cases, it is not clear how, since different regimes are defined by the cumbersome comparison between a comoving scale $k$ and a physical one $E_*$.

In order to respect the condition \Eq{valal} while avoiding unnaturally large values of $\a_0$, which would be difficult to justify theoretically, we consider only $|\a|=O(1)$. Another reason to do so is that the effective tensor running $\a^2\a_{\rm t}$ strongly suppresses the high-frequency spectrum if $\a=O(10)$. Therefore, this corner of the parameter space would not be verifiable anyway.

\begin{figure}
\centering
\includegraphics[width=10cm]{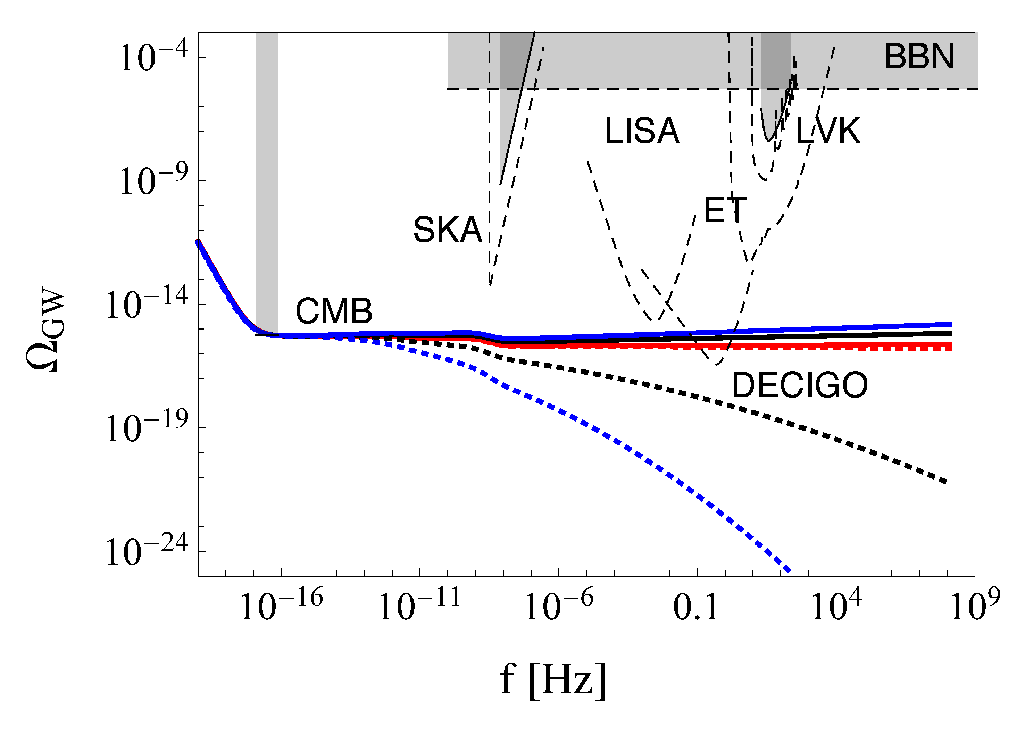}
\caption{\label{fig11} SGWB of multi-fractional inflation with no running ($\a_{\rm t}=0$) and $\a=-1/2,-3,-5$ (respectively, red, black and blue solid curve), compared with the sensitivity curves of  LIGO-Virgo-KAGRA (LVK), SKA, LISA, ET and DECIGO. Here $r=0.06$ and $n_{\rm t}\approx -0.0075$ is given by the consistency relation \Eq{r8n} and we take $k_*=10^{-3}\,{\rm Mpc}^{-1}$. The dotted curves correspond to the above cases with non-zero running $\a_{\rm t}=-0.0001$ (typical order of magnitude of inflationary models). Note that we widened the single-frequency CMB constraint to a band in order to make it more visible and that the spectra in the plot satisfy the CMB bound at the pivot scale.}
\end{figure}

As one can see in Fig.\ \ref{fig11}, for the typical signs and values of $r$, $n_{\rm t}$ and $\a_{\rm t}$ of standard inflationary scalar-field models, the theory can reach the DECIGO sensitivity curve if the running effect is negligible and the tensor-to-scalar ratio $r$ is close to the CMB bound (in the figure, $r=0.06$). The running decreases the spectral amplitude at the typical frequencies of interferometers. We see that the effect of running strongly depends on the value of $\a$, since the negative running is amplified by $\a^2$. Here we take $k_*=10^{-3}\,{\rm Mpc}^{-1}$, above which scale $\a$ is not constrained by the CMB observables of the scalar sector \cite{frc14}. Note that the results with zero running are not very sensitive to the choice of scale $k_*$ (position of the bending point of the spectrum) because $\a n_{\rm t}$ is very small, while the running effect strongly depends on the value of $k_*$. In the figure, we show the possible largest effect of the fractional regime by taking the minimally allowed value of $k_*=10^{-3}\,{\rm Mpc}^{-1}$, while the large suppression can be avoided if we take a larger value of $k_*$.

%%%%%%%%%%%%%%%%%%%%%%%%%%%%%%%%%%%%%%%%%%%%%%%%%%%%%%%%%%%%%%%%%%%%%%%%%%%%%%%%%%%%%%%%%%%%%%%%%%%%
%%%%%%%%%%%%%%%%%%%%%%%%%%%%%%%%%%%%%%%%%%%%%%%%%%%%%%%%%%%%%%%%%%%%%%%%%%%%%%%%%%%%%%%%%%%%%%%%%%%%

\section{The BBN bound}\label{sec5}

In the previous section, we have seen cases where a power law is not a good approximation to estimate the amplitude of the SGWB at high frequencies. However, the parametrization with $n_t$, $\a_t$ and so on is still useful to make a first rough prediction from the theory. Here we make a general discussion on the bound from the big-bang nucleosynthesis, which is the ultimate exclusion factor for models where the SGWB grows indefinitely at high frequencies. For example, in Fig.~\ref{fig8}, we find a blue-tilted SGBW intersect the BBN exclusion region in the upper right corner. Avoidance of this region places strong bounds on the tensor spectrum and on the underlying model of cosmic seeds, let it be inflation or alternatives such as cosmic-string cosmology or the ekpyrotic universe \cite{Stewart:2007fu,1792150}. In this section, we will show that the upper bound on the tensor index found in \cite{Stewart:2007fu} becomes stronger in the presence of positive running. We will also apply this result to the models considered above.
%and see how they can avoid the BBN bound while giving rise to a detectable signal.

%%%%%%%%%%%%%%%%%%%%%%%%%%%%%%%%%%%%%%%%%%%%%%%%%%%%%%%%%%%%%%%%%%%%%%%%%%%%%%%%%%%%%%%%%%%%%%%%%%%%

\subsection{BBN bound with tensor running}\label{bbnbo}

The constraint on the number of relativistic degrees of freedom during BBN imposes the upper limit
\be\label{intOm}
\int_{f_{\rm BBN}}^{f_{\rm e}}\frac{\rmd f}{f}\,\Om_\textsc{gw}(f)\,h^2<5.6\times10^{-6}\Delta N_\nu\,,
\ee
where $f=k/(2\pi)$, the frequency $f_{\rm BBN}$ corresponds to the Hubble rate at BBN, $f_{\rm e}$ is a sharp cut-off set by the Hubble rate $H_{\rm e}$ at the end of the period generating cosmic perturbations (inflation or alternatives) and $\Delta N_\nu$ parametrizes the effective number of extra relativistic degrees of freedom at the time of BBN. The current upper bound is $\Delta N_\nu < 0.41$ (BBN+$Y_p$+D, 95\% CL) \cite{Cyburt:2015mya}. A common procedure is to ignore the logarithmic integral, which is tantamount to considering the dominant contribution from the peak of the spectrum. This gives the constraint on high-frequency ($f > f_{\rm BBN}$) GWs as 
\be\label{intOm2}
\Om_\textsc{gw}(f)\,h^2<2.3\times10^{-6}.
\ee
For a mode entering the horizon during the radiation-dominated phase with $g_\ast=g_{\ast s}=106.75$, the transfer function \Eq{Tk} becomes 
\be
\label{Tk2}
\cT^2 (k, \tau_0) \simeq 0.37\frac{\Omega_{\rm m}^2 H_0^4}{k^2 k_{\rm eq}^2}\,.
\ee 
Substituting \Eq{Ptk} and \Eq{Tk2} into \Eq{Omgw}, omitting the dependence on the pivot scale $f_0$ in the observables $r$, $n_{\rm t}$ and $\a_{\rm t}$, and plugging the \textsc{Planck} bound \Eq{psbo}, we can approximate the spectral shape as
\ba
\Om_\textsc{gw}(f) &=&\frac{k^2}{12H_0^2}\cT^2(f)\,\cP_{\rm t}(f)
\simeq 0.031\frac{\Omega_{\rm m}^2 H_0^2}{k_{\rm eq}^2} r\,\cP_{\rm s}(f_0) \left(\frac{f}{f_0}\right)^{n_{\rm t}+\frac{\a_{\rm t}}{2}\ln\frac{f}{f_0}}\nonumber\\
&=:&\frac{\cA}{h^2}\,r\left(\frac{f}{f_0}\right)^{n_{\rm t}+\frac{\a_{\rm t}}{2}\ln\frac{f}{f_0}},\qquad \cA\approx 1.4\times 10^{-15} 
\label{Omgw2}\,,
\ea
which updates the estimate on $\cA$ of \cite{Chongchitnan:2006pe} based on WMAP data and the transfer function of \cite{Turner:1993vb}. One can plug \Eq{Omgw2} into \Eq{intOm} and integrate exactly, but we checked that the constraints below do not change for any value of $\a_{\rm t}>0$ if instead we use \Eq{intOm2}, so that
\be\label{boundOm0}
\cA\,r\left(\frac{f}{f_0}\right)^{n_{\rm t}+\frac{\a_{\rm t}}{2}\ln\frac{f}{f_0}}<2.3\times10^{-6}\,.
\ee
In the limit $\a_{\rm t}\to 0$, this reduces to the expression of \cite{Stewart:2007fu}. Otherwise, we get
\be\label{boundOm}
n_{\rm t}<\frac{\ln\left(\frac{2.3\times10^{-6}}{\cA\,r}\right)}{\ln\frac{f}{f_0}}-\frac{\a_{\rm t}}{2}\ln\frac{f}{f_0}\,.
\ee
Now we can use the bound \Eq{boundOm} to place a constraint on the tensor index $n_{\rm t}$. The parameter $\cA$ was given in \Eq{Omgw2} and we set $r=0.068$ to its upper bound \Eq{boundr}. The pivot scale \Eq{pivot} and the BBN scale correspond, respectively, to
\be
f_0= \frac{k_0}{2\pi}\frac{\lpl}{\tpl}\approx 7.73\times 10^{-17}\,{\rm Hz}\,,\qquad f_{\rm BBN}= \frac{a_{\rm BBN} H_{\rm BBN}}{2\pi}\frac{\mpl}{\tpl}\approx 10^{-10}\,{\rm Hz}\,,
\ee
where we multiplied by Planck units conversion factors $\lpl/\tpl\approx9.72\times 10^{-15}\,{\rm Mpc}\,{\rm s}^{-1}$ and $\mpl/\tpl\approx 2.26\times 10^{62}\,{\rm GeV}\,{\rm s}^{-1}$, respectively. For the high-frequency cut-off $f_{\rm e}$, we may choose the comoving Planck scale $a_{\rm e}/\tpl$ for string-gas cosmology and the new ekpyrotic scenario, while for inflationary models it is determined by the Hubble scale at the end of inflation, $H_{\rm e}$. For simplicity, 
we calculate it for instant reheating, in which case the temperature of the universe at the end of inflation or alternative models is given by 
\be
T_{\rm e} = \left[\frac{90}{\pi^2g_*(T_{\rm e})}\right]^{1/4} \left(\Mpl H_{\rm e}\right)^{1/2}\,,
\ee
where 
%$g_*(T_{\rm e})=106.75$ is 
we take $g_*(T_{\rm e})=106.75$ for the number of relativistic degrees of freedom at the end of inflation. Then the scale factor corresponding to $T_{\rm e}$ is 
\be
\frac{a_{\rm e}}{a_0}=\left[\frac{11}{43}g_*(T_{\rm e})\right]^{-1/3}\frac{T_0}{T_{\rm e}}\,,
\ee
where $T_0=2.725 {\rm K} \approx 2.35 \times 10^{-13}\,{\rm GeV}$ is the temperature of the universe today. Setting $T_{\rm e}=10^{16}\,{\rm GeV}$, we get the expansion factor $a_{\rm e}\approx 7.80\times 10^{-30}$ at the grand-unification scale. For string-gas cosmology and the new ekpyrotic scenario, the cut-off scale is the Planck scale dilated up to grand-unification, so that and we find that the high-frequency cut-off is given by 
\be
f_{\rm e}=f_{\rm Pl}=\frac{a_{\rm e}}{2\pi\tpl}\approx 2.30\times 10^{13}\,{\rm Hz}\,,
\ee
while
\be
f_{\rm e}=f_{\rm inf}=\frac{a_{\rm e} H_{\rm e}}{2\pi}\approx 2.65\times 10^{8}\,{\rm Hz}\,,
\ee
for grand-unification scale inflation. With these values, in the absence of running 
% \be\label{upbo}
% n_{\rm t}< 0.33\qquad (\a_{\rm t}=0)\,,
% \ee
%which increases to $n_{\rm t}< 0.38$ when $f_{\rm e}=f_{\rm inf}=a_{\rm e} H_{\rm e}/(2\pi)\approx10^{-4}f_{\rm Pl}$ is the grand-unification scale typical of large-field inflation. 
we find $n_{\rm t}< 0.35$ and $n_{\rm t}< 0.42$ respectively for $f_{\rm e}=f_{\rm Pl}$ and $f_{\rm e}=f_{\rm inf}$ with $r=0.068$. Thus, taking the largest energy cut-off gives the tightest constraint on $n_{\rm t}$.

\begin{figure}
\centering
\includegraphics[width=10cm]{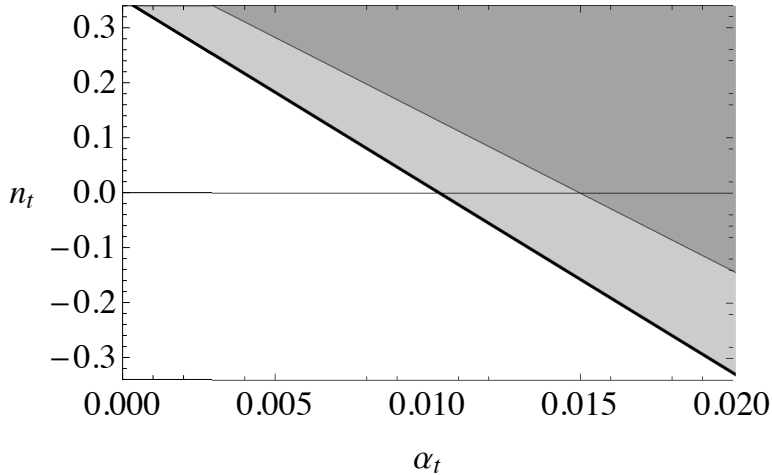}
\caption{\label{fig12} Upper bound \Eq{boundOm} on the tensor spectral index as a function of the running $\a_{\rm t}$, with the frequency and spectral values given in the text, for $f_{\rm e}=f_{\rm Pl}$ (thick line) and $f_{\rm e}=f_{\rm inf}$ (thin line). We fix the tensor-to-scalar ratio to be $r=0.068$ and assume that inflation or its alternative model occurs at the grand-unification scale, i.e. $T_e=10^{16}$\,GeV. The exclusion region is shaded.}
\end{figure}

When $\a_{\rm t}\neq0$, the bound \Eq{boundOm} is 
%relaxed if the running is negative and strengthened if it is positive
strengthened if the running is positive (note that $\ln f_{\rm e}/f_0\approx 70$ is positive). 
To show the effect of running on the BBN bound, we plot this bound as a function of $\a_{\rm t}>0$ in Fig.\ \ref{fig12}. We find that the inclusion of running is important since it dramatically changes the constraint on $n_t$. This is because it is much easier to get a large signal at high frequencies when we have positive $\a_t$, since the effect is enhanced by a factor of $\ln f_{\rm e}/f_0\approx 70$ compared to $n_t$, as seen in \eqref{Omgw2}. When $\a_{\rm t}\gtrsim 0.01$, even a red-tilted tensor spectrum can saturate the bound. 
%Thus, the larger the tensor running, the easier to get a large signal at the frequencies of interest.

%%%%%%%%%%%%%%%%%%%%%%%%%%%%%%%%%%%%%%%%%%%%%%%%%%%%%%%%%%%%%%%%%%%%%%%%%%%%%%%%%%%%%%%%%%%%%%%%%%%%

\subsection{Application to the models}

%The result \Eq{boundOm} given in Fig.\ \ref{fig12} should be applied to the models of quantum gravity considered in this paper. 
Let us consider applications of the BBN bound to the five models discussed in section \ref{sec4}.

%As we have seen, even when $n_{\rm t}=O(10^{-2})$ is well below the upper bound \Eq{upbo}, the tensor running can make the spectrum rise up to the BBN threshold at some high frequency. 
In the case of models based on scalar-field inflation, i.e, non-local Starobinsky inflation, Brandenberger--Ho non-commutative inflation and multi-fractional inflation, we have a weaker bound from BBN compared to the cases of string-gas cosmology and the new ekpyrotic scenario. This happens because the high-frequency cut-off is at the lower frequency $f_{\rm inf} \ll f_{\rm Pl}$. We have seen that all the former three models do not predict values of $n_{\rm t}$ and $\a_{\rm t}$ large enough to reach the BBN bound, so that these scenarios are safe.

In the case of string-gas cosmology and the new ekpyrotic scenario, a SGWB could be produced up to the frequency corresponding to the Planck scale, so that the BBN bound is stronger. For string-gas cosmology, we have shown that there is a theoretical upper bound on the GW amplitude at high frequencies. This upper bound goes beyond the BBN line only when $l_{\rm st} \lesssim 4.5\,\Mpl^{-1}$ which is already excluded by the CMB bound. Thus, for this model, the BBN bound does not provide further constraints. For the new ekpyrotic scenario, we have found that the SGWB amplitude can be large at high frequencies if $\a_{\rm t} = -\a_{\rm s}$ takes a positive value. By taking $f_{\rm e}=f_{\rm Pl}$, we find that $\a_{\rm s} > -0.012$ should be satisfied in order not to exceed the BBN bound (this constraint does not depend much on the choice of $n_{\rm t}= 1-n_{\rm s}$, since the range of $n_{\rm s}$ is narrow and does not affect the GW amplitude much, as shown in Fig.~\ref{fig8}). If we respect this constraint, the SGWB amplitude does not reach the sensitivity curve of ET.

Note that the BBN bound %\Eq{upbo} and its generalization \Eq{boundOm} to a non-zero running 
 can be avoided if the inflationary phase (or the epoch of perturbations generation by an alternative mechanism) ends with reheating and the evolution of the universe after that follows general relativity. In fact, in that case the tensor spectrum bends down as $\sim k^{n_{\rm t}-2}$ \cite{Seto:2003kc}, thus evading the BBN constraint \cite{Kuroyanagi:2014nba,Kuroyanagi:2020sfw}. The position of the bending point depends on the reheating temperature $T_{\rm reh}$, ranging from 
%$f\sim 10^{-2}\,{\rm Hz}$ ($T_{\rm reh}=10^6\,{\rm GeV}$) 
$f\sim 10^{-10}\,{\rm Hz}$ ($T_{\rm reh}=10^{-2}\,{\rm GeV}$, the lowest reheating temperature preserving BBN) to $f\sim 10^8\,{\rm Hz}$ ($T_{\rm reh}=10^{15}\mhyp 10^{16}\,{\rm GeV}$, instantaneous reheating).

\section{Discussion and conclusions}\label{sec6}

To summarize:
\begin{itemize}
\item \emph{Non-local Starobinsky inflation}. This model is out of reach of any present or future GW experiment because at high frequencies it reduces to standard Starobinsky inflation and, afterwards, to the standard cosmic evolution.
\item \emph{String-gas cosmology}. Depending on the value of the scalar index and running, which determine their tensor counterparts, the SGWB can reach the DECIGO sensitivity range.
\item \emph{New ekpyrotic scenario}. The SGWB reaches the ET and DECIGO sensitivity curve only in the most optimistic choice of parameters. However, if we assume that the spectrum grows continuously towards high frequencies, with no reheating mechanism, then the BBN bound excludes the parameter space accessible by ET.
\item \emph{Brandenberger--Ho non-commutative inflation}. Since this model deviates from standard inflation only in the sign and value of the coefficients in the slow-roll expansions of the observables, the tensor spectrum is blue-tilted but the effect is weak. The SGWB reaches DECIGO sensitivity but it does not go beyond that.
\item \emph{Multi-fractional inflation}. If the tensor-to-scalar ratio is large enough, but still below the observational upper bound \Eq{boundr}, then the SGWB can reach DECIGO sensitivity already for natural small values of the fractional parameter $\a<0$, for which the negative tensor running is negligible. The signal increases with increasing $|\a|$ but, at the same time, so does the suppression effect of the running.
\end{itemize}

For good or for worse, these results are affected by the running of the tensor index $\a_{\rm t}$, which in many cases turns out to be large enough at high frequencies to increase or to lower the tilt up to or below the sensitivity curves of the experiments. Therefore, this often-ignored inflationary observable deserves more attention in quantum gravity, where quantum corrections can compensate the small higher-order slow-roll coefficients in the observables. 

However, we also have seen cases where the power-law series expansion, parametrized by $n_{\rm t}$, $\a_{\rm t}$ and higher-order slow-roll observables, does not correctly describe the GW spectrum at interferometer scales. For non-local Starobinsky inflation, we compared the spectra calculated using the approximated form and the exact form, and found that even the inclusion of higher-order terms does not help to get the correct prediction of the GW amplitude. This is because, for Kuz'min and Tomboulis form factors, the higher-order derivatives of the quantum-gravity correction term diverges in the regime where the non-local effect is large. For the case of Brandenberger--Ho non-commutative inflation and multi-fractional inflation, we found that corrections are small in general, so that these models preserve a hierarchy of slow-roll parameters and observables and the standard power-law series expansion is a good approximation for all the frequencies of interest. The cases of string-gas cosmology and the new ekpyrotic scenario are more delicate because they are not inflationary models and the evolution of the universe towards the reheating phase has not been investigated in detail. They could even show a non-trivial behaviour at high frequencies which cannot be parametrized by slow-roll observables. Our exploration of quantum-gravity theories with blue-tilted tensor spectrum does not pretend to be exhaustive and there may be other models with such a feature worth studying (e.g., \cite{Dapor:2020jvc}). One such example is pre-big-bang scenario \cite{Gasperini:1992em}, where $r<0.01$ at the pivot scale \Eq{pivot} and the SGWB is blue-tilted and can reach the sensitivity of present and future interferometers \cite{Gasperini:2016gre}.
%Also, except for non-local Starobinsky inflation, we have not considered the role of the running of the running $\rmd\a_{\rm t}/\rmd\ln k$ due to the strong slow-roll suppression but it may be worthy of investigation in the future. We do not foresee any impact in the case of Brandenberger--Ho non-commutative inflation and multi-fractional inflation, both having a hierarchy of slow-roll parameters and observables and respecting the standard slow-roll approximation. However, in the case of string-gas cosmology and the ekpyrotic scenario the tensor index and running are not fully determined by the theory but by consistency relations constrained by observations, so that we do not known whether higher-order observables (running of the running and so on) are really suppressed with respect to the first elements of the hierarchy; see Tab.\ \ref{tab1}. In this paper, we implicitly assumed so when we adopted the parametrization \Eq{Ptk} for the tensor spectrum.

%Finally, evasion of the BBN upper bound on the SGWB amplitude is guaranteed one way or another, as discussed in the previous section, but it would deserve to be explored more in detail. 
Our detailed investigation on the representative five models in section~\ref{sec4} have shown that the commonly used power-law series parametrization of the primordial tensor spectrum is not always appropriate and it must be superseded by a case-by-case analysis using the full spectrum. Furthermore, constraints on the models by the BBN upper bound (and by interferometer experiments in the future) could strongly depend not only on the high-frequency behaviour of the primordial spectrum, but also, as briefly mentioned in section~\ref{sec5}, on the reheating mechanism following inflation or its alternatives, which would deserve to be explored more in detail. All these features can enrich the quest for the cosmological imprint of quantum gravity.

%%%%%%%%%%%%%%%%%%%%%%%%%%%%%%%%%%%%%%%%%%%%%%%%%%%%%%%%%%%%%%%%%%%%%%%%%%%%%%%%%%%%%%%%%
%%%%%%%%%%%%%%%%%%%%%%%%%%%%%%%%%%%%%%%%%%%%%%%%%%%%%%%%%%%%%%%%%%%%%%%%%%%%%%%%%%%%%%%%%

\section*{Acknowledgments}

\noindent The authors thank R.\ Brandenberger for useful correspondence. They are supported by the I+D grant FIS2017-86497-C2-2-P of the Spanish Ministry of Science and Innovation and acknowledge networking support by the COST Action CA18108. S.K.\ is supported by the Atracción de Talento contract no.\ 2019-T1/TIC-13177 granted by the Comunidad de Madrid in Spain, and by the Japan Society for the Promotion of Science (JSPS) KAKENHI Grants no.\ 20H01899 and 20J40022.

%%%%%%%%%%%%%%%%%%%%%%%%%%%%%%%%%%%%%%%%%%%%%%%%%%%%%%%%%%%%%%%%%%%%%%%%%%%%%%%%%%%%%%%%%
%%%%%%%%%%%%%%%%%%%%%%%%%%%%%%%%%%%%%%%%%%%%%%%%%%%%%%%%%%%%%%%%%%%%%%%%%%%%%%%%%%%%%%%%%

\end{document}